\documentclass[10pt,onecolumn,amsmath,amssymb]{article}
\usepackage{amsfonts}
\usepackage{amsmath}
\usepackage{amssymb}
\usepackage{graphicx}
\usepackage{bm}
\usepackage[super]{cite}
\usepackage{times}

\usepackage[font={small}]{caption}
\usepackage[title,titletoc]{appendix}

\newcommand{\com}[1]{}

\newcommand{\onlinecite}[1]{\citen{#1}}

\newcommand{\V}[1]{|{#1}\rangle}
\newcommand{\iV}[1]{\langle{#1}|}
\newcommand{\aV}[1]{\langle{#1}\rangle}

\newcommand{\MxQ}[2]{
\begin{pmatrix} 
#1 \\ 
#2 
\end{pmatrix}}
\newcommand{\MxQQ}[4]{
\begin{pmatrix} 
#1\\ 
#2\\ 
#3\\ 
#4
\end{pmatrix}}

\newcommand{\MxRRR}[3]{
\begin{pmatrix} 
#1 \\ 
#2 \\ 
#3 
\end{pmatrix}}

\makeatletter
\renewcommand\section{\@startsection{section}{1}{\z@}%
{-3.5ex \@plus -1ex \@minus -.2ex}%
{2.3ex \@plus.2ex}%
{\normalfont\bfseries}}
\renewcommand\subsection{\@startsection{subsection}{1}{\z@}%
{-3.5ex \@plus -1ex \@minus -.2ex}%
{2.3ex \@plus.2ex}%
{\normalfont\bfseries}}
\makeatother

\date{}

\begin{document}

\title{\large\bf
Quantum walks as mathematical foundation for quantum gates
}
\author{
\\\normalsize\it
Dmitry Solenov
\\\normalsize\it
Department of Physics, St. Louis University, St. Louis, Missouri  63103, USA
\\\normalsize\it
solenov@slu.edu
}

\maketitle\thispagestyle{empty}

\begin{abstract}
\noindent
It is demonstrated that in gate-based quantum computing architectures quantum walk is a natural mathematical description of quantum gates. It originates from field-matter interaction driving the system, but is not attached to specific qubit designs and can be formulated for very general field-matter interactions. It is shown that, most generally, gates are described by a set of coined quantum walks. Rotating wave and resonant approximations for field-matter interaction simplify the walks, factorizing the coin, and leading to pure continuous time quantum walk description. The walks reside on a graph formed by the Hilbert space of all involved qubits and auxiliary states, if present. Physical interactions between different parts of the system necessary to propagate entanglement through such graph---quantum network---enter via reduction of symmetries in graph edges. Description for several single- and two-qubit gates are given as examples.

\end{abstract}

\tableofcontents


\section{Introduction}\label{sec:I}

Quantum information and computing relies on few basic quantum mechanical concepts, such as quantum state, quantum superposition, quantum entanglement, quantum measurement. In it's standard form---gate-based quantum computing---it relies on quantum gates \cite{DiVincenzo,Barenco}, manipulating superpositions and entanglement. Supplemented by quantum measurements, it can process information and solve complex problems at the rate not accessible to classical information processing.\cite{nielsenchuang} As such, quantum computing is one of the best tests of basic quantum mechanical principles abstracted out from actual physical systems implementing it. Yet, in gate-based architectures, quantum gates have been intimately connected with physical systems in which they are implemented or for which they are designed.\cite{Chow,solenov1,solenov2,solenov3,solenov4} Furthermore, quantum gate designs has been closely following developments in physical qubit architectures.\cite{Koch,Awschalom,Wrachtrup,Paik,Rigetti,Nigg,Blinov,CiracZoller} On the other hand, substantial effort has been ongoing to optimize quantum algorithms to make them run faster\cite{Markov,Saeedi-1,Saeedi-2} and correct errors\cite{Aharonov,Shor,Kitaev,Bell} to reach fault tolerance. This effort however is largely confined to elementary quantum gates supplied by specific qubit designs, with the only degree of freedom being arrangement of such quantum gates. 
The combination of such two approaches largely mimics current classical information processing strategy leaning towards RISC (Reduced Instruction Set Computing) CPU architectures. This is not necessarily an optimal way of using physical resources available in quantum computing systems, which was also the case for classical computing, benefiting from CISC (Complex Instruction Set Computing) approach at early stages.  

Quantum principles are unique as they incorporate freedom that is not present in classical deterministic description---quantum particles are free to evolve in provided Hilbert space until measured. This situation is in the very core of quantum computing concept. Yet, design of quantum gates and quantum code compression typically aims to suppress this freedom by relying on deterministic sequences of controls micromanaging quantum evolution and trajectories. Recent analysis of few-qubit (entangling) quantum gates performed via continuous time quantum walks driven by classical field \cite{Solenov-QW} have shown that by allowing greater freedom for quantum particles during multiqubit rotations one can significantly speedup entangling quantum gates. Yet, formulation of such gates have still been attached to chosen qubit architectures by relying on resonant approximation that is very specific to the actual physical system at hand.

Here I show that quantum walks\cite{Kempe,Andraca,Kendon,KendonTregenna,Fedichkin1,Fedichkin2,Tamon} is a general framework for quantum gates that is not tight to a specific physical realization or qubit architecture, as far as I focus on gate-based quantum computing strategy. I demonstrate that quantum gates are, in general, described by a collection of {\it coined} quantum walks realized in a quantum network of available (multiqubit) quantum states. Physical interactions necessary to carry and propagate entanglement enter via symmetry of edges (connections) independently of whether resonant approximation is used. Typical approximations, such as rotating wave\cite{van-Kampen,Mandel-Wolf,Abragam} and resonant\cite{solenov4,Solenov-QW} approximations, are naturally described withing the quantum walk approach and can be verified by specifically designed quantum walks. I show that when resonant approximation is appropriate, quantum coins are factored out and quantum gates are described by continuous time quantum walks. In the latter case quantum walks constructed to implement quantum gates bear some similarity with walks used in quantum-wire-based architectures\cite{Childs}. In contrast to the quantum wire architectures, quantum walks constructed to implement gates are naturally controlled by classical time-dependent control field (pulses) via graph edges. As the result, this description incorporates standard quantum gates schemes when the size of available quantum network is reduced to a minimum and multiple pulses are used instead of one. In the rest of this section I briefly introduce the concepts of coined quantum walks, continuous time quantum walks, and quantum gates as needed for subsequent sections. Quantum walks description is derived in Sec.~\ref{sec:QWG} and summarized in Sec.~\ref{sec:Gs} where it is used to formulate quantum gates. Sections~\ref{sec:QWG-1D} and \ref{sec:QWG-GEN} show how some walks can be reduced to one-dimensional walks, which are easier to solve, and generalize results to more complex filed-matter interactions.

\subsection{(coined) quantum walks}\label{sec:I:QW}

Quantum walks are typically introduced by analogy with classical random walks.\cite{Kempe,Andraca,Kendon} Considering walk on an infinite line as an example, one can define an amplitude of shifting to the left adjacent site or to the right adjacent site. In this case the wave function of a walking particle, initially localized at site ``0'' is
\begin{eqnarray}\label{eq:QW:psi-1D}
\V{\psi(t+\Delta t)} = A \V{-1} + B \V{+1},
\quad\quad
|A|^2 + |B|^2 = 1,
\end{eqnarray}
where $\Delta t$ time interval counts steps. One way to achieve this starting with state $\V{0}$ is to use the state of a qubit (two-sate quantum system), typically referred to as ``quantum coin,'' to supply amplitudes for the two different directions 
\begin{eqnarray}\label{eq:QW:psi-1D-coin}
\V{\psi(t+\Delta t)} = A \V{-1,0} + B \V{+1,1}.
\end{eqnarray}
Here the second state index denotes the basis states of the coin. The coin amplitudes can be superimposed onto the shifted states of the walking particle by a shift operator $\hat{S}$ that couples the two systems. The state of the coin can be rotated prior to applying $\hat{S}$ by a coin step operator $\hat{C}$ to influence the walk. Such quantum walks are  referred to as coined quantum walks. Although the term ``discrete time'' quantum walk is used as well. Note that the notion of time here is not critical as it enters only via a sequence of events (steps), while the actual value of $\Delta t$ is irrelevant and can be set to $\Delta t\to dt\to 0$, if needed. Both, shift and coin operators are unitary rotations
\begin{eqnarray}\label{eq:QW:S-C-unitary}
\hat{S}\hat{S}^\dag = 1,
\quad\quad
\hat{C}\hat{C}^\dag = 1.
\end{eqnarray}
The overall step operator is
\begin{eqnarray}\label{eq:QW:psi}
\V{\psi(t+\Delta t)} = \hat{S} \hat{C} \V{\psi(t)}
\end{eqnarray}
In the case of a walk on a line, the shift operator can be defined as
\begin{eqnarray}\label{eq:QW:S}
\hat{S}\V{\xi,c} = \V{\xi + (-1)^c,c}
\quad
\to
\quad
\hat{S} = 
\sum_\xi\V{\xi+1}\iV{\xi}\otimes \V{0_c}\iV{0_c}
+
\sum_\xi\V{\xi-1}\iV{\xi}\otimes \V{1_c}\iV{1_c},
\end{eqnarray}
where indexes $\xi$ denote vertices on the line (states) the walker can occupy, and $c=0,1$ refers to the basis states of the coin (qubit). We will omit the direct product from now on and will use different state operators or indices to refer to the particle they operate on. In general, the shift operator can be any (unitary) operator that couples to the state of the coin and move the walker, e.g., a (more natural) quantum evolution operator
\begin{eqnarray}\label{eq:QW:S-exp}
\hat{S} = e^{-i\gamma
\sum_\xi\left(
\V{\xi+1}\iV{\xi}\V{0_c}\iV{1_c} + h.c.
\right)
},
\end{eqnarray}
where $\gamma$ is some real number and $h.c.$ stands for hermitian conjugate terms. The specific operator chosen will implement specific walk. The coin does not have to be a two-state system---it can have as many states as needed to determine the walk on available network (graph) of states. Note that if the number of connections in the graph varies from vertex to vertex, the coin must be local to accommodate this change. Similarly, $\hat{C}$ is any unitary operator, such as, e.g., a Hadamard operator
\begin{eqnarray}\label{eq:QW:C}
\hat{C} = \hat{\rm H}\equiv \frac{1}{\sqrt{2}} \MxQ{1 & 1}{1 & -1}
\end{eqnarray}
in the case of a two-state coin. Because the system involves two quantum objects and we, typically, do not have access to the state of the coin, the state of the worker is analyzed by looking at the reduced density matrix with the coin degrees of freedom traced out
\begin{eqnarray}\label{eq:QW:rho}
\hat{\rho}(t) = {\rm Tr}_c(\V{\psi(t)}\iV{\psi(t)}) \equiv \sum_{c=0,1}\aV{c|\psi(t)}\aV{\psi(t)|c}.
\end{eqnarray}
Note that, in general, the walker becomes entangled with the coin and the density matrix $\rho(t)$ does not correspond to any pure state
\begin{eqnarray}\label{eq:QW:rho-not-psi}
\hat{\rho}(t) \neq \V{\psi}\iV{\psi}
\end{eqnarray}
The probability to find the walking particle at site $\xi$ can be easily found as
\begin{eqnarray}\label{eq:QW:P}
P_\xi(t) = \iV{\xi}\hat{\rho}(t)\V{\xi}
\end{eqnarray}

\subsection{continuous time quantum walks}\label{sec:I:CTQW}

Continuous time quantum walks do not rely on auxiliary quantum coins to propagate. It is an evolution due to a unitary rotation\cite{Kendon} 
\begin{eqnarray}\label{eq:CTQW:psi}
\psi(t+dt) = e^{-idt\hat{H}} \psi(t)
\end{eqnarray}
defined by Hamiltonian, $\hat{H}$, and, thus, is simply a wave function of a discrete-state quantum system evolving according to the Schroedinger equation 
\begin{eqnarray}\label{eq:CTQW:SCE}
i\frac{d}{dt}\psi(t) = \hat{H} \psi(t).
\end{eqnarray}
Here and in what follows we will use the same units for energy and frequency, in which case $\hbar = 1$.
Continuous time quantum walks are typically based on time-independent Hamiltonians with all dimensional parameters often lumped together into a prefactor
\begin{eqnarray}\label{eq:CTQW:A}
\hat{H} = \gamma \hat{A},
\end{eqnarray}
although continuous time (classical) random walks on time-dependent graphs are possible and so is the quantum analogy with $\hat{H}(t)$. The remaining matrix $\hat{A}$ is adjacency matrix describing connections in the given graph. In graph theory adjacency matrices with only 0 or 1 entries are natural. In this case 1 denotes existing connection (edge) between the two vertices (column and row indexes) and 0 denotes no connection. A physical process corresponding to some Hamiltonian and incorporating several parameters generally includes complex-number entries in $\hat{A}$ and factorization of all parameters out of the matrix structure is not common, although possible in some cases. Diagonal entires of the Hamiltonian (in a given basis) have the meaning of energies of the basis states. They can be incorporated into the adjacency matrix as self-loops (connecting a vertex to itself).

Continuous time and coined quantum walks are substantially different. Coined walk incorporates additional quantum object (coin) interacting with the walker. Yet they can result in an identical evolution in some special cases when the coin can be factored out, so that the reduced density matrix (\ref{eq:QW:rho}) remains the outer product of $\V{\psi}$ and its hermitian conjugate. Both types of quantum walks were proven to provide similar quantum speedup as regular quantum computing.\cite{Childs,Spielman,Lovett}

Recently it was demonstrated \cite{Solenov-QW} that continuous time quantum walks is a natural description of hardware quantum gates taking advantage of extended Hilbert space available in many qubit architectures targeting gate-based quantum computing. A collection of auxiliary states (states beyond the boolean computational domain), many of which participate in cross-qubit interactions, can be naturally viewed as non-boolean (not qubit-based) quantum network. Appropriately designed continuous time quantum walks through such networks accumulate nontrivial phase faster then traditional entangling gates under the same condition but relying on only small part of such network each time. However these earlier derivations\cite{Solenov-QW} were obtained in resonant and rotating wave approximations. The question of whether quantum walk description is valid in a more general case, when these approximations are inappropriate, was not resolved. 

In this work it is demonstrated that quantum walk is a natural mathematical framework to construct quantum gates in qubit systems (with or without auxiliary states) controlled via classical field (pulses). In the case when non-resonant physics can not be ignored, quantum gates utilizing quantum networks are described by coined quantum walks. They turn into continuous time quantum walks (without auxiliary quantum coins) when only resonant processes are relevant. Both types of walks are based on the same graph with non-resonant processes entering primarily via phases introduced by a feedback via the coin (or coins in multi-mode case). In order to demonstrate this we first briefly discuss quantum gates and show why continuous time quantum walks can potentially emerge as a description.

\subsection{quantum gates}\label{sec:QG}

In gate-based quantum computing, a quantum gate is a coherent rotation of the wave function by any unitary operator
\begin{eqnarray}\label{eq:QG:Ug}
\V{\Psi'} = \hat{U}_g \V{\Psi}.
\end{eqnarray}
Because wave functions are normalized superpositions of given basis states with some complex amplitudes
\begin{eqnarray}\label{eq:QG:psi}
\V{\Psi} = \sum_{\xi_1,\xi_2,...} B_{\xi_1,\xi_2,...} \V{\xi_1\xi_2...},
\end{eqnarray}
the gate, in fact, is the rotation of the basis in which the wave function is considered (or constructed)
\begin{eqnarray}\label{eq:QG:Ug-psi-basis}
\sum_{\xi_1,\xi_2,...} B_{\xi_1,\xi_2,...} \V{\xi_1\xi_2...}' 
= 
\sum_{\xi_1,\xi_2,...} B_{\xi_1,\xi_2,...} \hat{U}_g\V{\xi_1\xi_2...}.
\end{eqnarray}

Entanglement---a basis dependent property of a multiqubit quantum system---can be altered by gates that perform global rotations of basis. Local (single-qubit) rotations, such as, e.g., Hadamard gate,
\begin{eqnarray}\label{eq:QG:Hadamard}
\frac{\V{0}+\V{1}}{\sqrt{2}} = \hat{U}_g({\rm H})\V{0},
\quad\quad
\frac{\V{0}-\V{1}}{\sqrt{2}} = \hat{U}_g({\rm H})\V{1},
\end{eqnarray}
do not change entanglement and only affect single-qubit superpositions for individual qubits or for many qubits at the same time if applied concurrently. Entangling gates, such as, e.g., a two-qubit control-NOT (CNOT) defined as
\begin{eqnarray}\label{eq:QG:CNOT}
\V{i,(j+i)\,{\rm mod}\,2} = \hat{U}_g({\rm CNOT,1})\V{i,j},
\\\label{eq:QG:NOTC}
\V{(i+j)\,{\rm mod}\,2,j} = \hat{U}_g({\rm CNOT,2})\V{i,j},
\end{eqnarray}
alter multi- (two-) qubit superpositions, but can not, in general, be applied concurrently if they share qubits.

In both cases rotations of basis states are equivalent to the end result of a set of continuous time quantum walks
\begin{eqnarray}\label{eq:QG:Ug-psi-basis}
\V{\xi_1\xi_2...}' = \hat{U}_g\V{\xi_1\xi_2...}
\end{eqnarray}
In order to produce a gate, all such walks must be coordinated to produce the results such as those stated in the above examples in Eqs.~(\ref{eq:QG:Hadamard}), (\ref{eq:QG:CNOT}), and (\ref{eq:QG:NOTC}). Equation~(\ref{eq:QG:Ug-psi-basis}), however, does not yet prove that quantum walk provide any additional insight into construction of quantum gates. To do so we must derive an evolution operator corresponding to a single step of each such walk based on physical description of control used to perform the gate on hardware qubits. This is done in the next section.

\section{Quantum walks as framework for quantum gates}\label{sec:QWG}

In gate-based quantum computing interaction between control apparatus and qubit system occurs via a classical control field, i.e., via a bosonic field defined by commutation relations between its creation/annihilation operators
\begin{eqnarray}\label{eq:QWG:a-a^dag}
[\hat{a},\hat{a}^\dag] = 1,
\end{eqnarray}
and characterized by the coherent state wave function\cite{Louisell}
\begin{eqnarray}\label{eq:QWG:a-coherent-states}
\V{\alpha} = e^{-|\alpha|^2/2}e^{\alpha \hat{a}^\dag}\V{0},
\quad
\hat{a}\V{\alpha} = \alpha\V{\alpha},
\quad
\iV{\alpha}\hat{a}^\dag = \alpha^*\iV{\alpha},
\quad
\aV{\alpha|\alpha'} = e^{
\alpha'\alpha^*-\frac{|\alpha|^2+|\alpha'|^2}{2},
}
\end{eqnarray}
where $\alpha$ and $\alpha^*$ are eigenvalues of annihilation and creation operators respectively. Such field becomes fully classical in the thermodynamic limit of large (average) number of bosons
\begin{eqnarray}
\label{eq:TD-limit}
N = \aV{\hat{a}^\dag \hat{a}} = |\alpha|^2 \to \infty.
\end{eqnarray}

A dipole field-matter interaction Hamiltonian is sufficient in most cases of qubit control (more general coupling is considered in the next section). In rotating frame (interaction representation), it can be written as
\begin{eqnarray}\label{eq:QWG:V}
\hat{V}(t) = 
\sum_p
\Phi_p(t) 
\!\!\!\!\!\!\!
\sum_{\omega_p;\,\vec{\xi}>\vec{\xi}'\in G}
\!\!\!\!\!\!\!
\left({\frak E}^*_{p,\omega_p}e^{i\omega_pt}+{\frak E}_{p,\omega_p}e^{-i\omega_pt}\right)
\left(
g^*_{\vec{\xi},\vec{\xi}'} 
\V{\vec{\xi}}\iV{\vec{\xi}'}
e^{-i \Delta E_{\vec{\xi},\vec{\xi}'} t}
+
h.c.
\right).
\end{eqnarray}
Here the system of qubits and, possibly, additional (auxiliary) sates is represented by a graph $G$ with vertices labeled by $\xi$, and the subgraph
\begin{eqnarray}\label{eq:QWG:G2}
G_Q \in G
\end{eqnarray}
representing all qubit states (qubit or boolean domain).
Each pulse of the control field (indexed by $p$) has the overall envelop profile $\Phi(t)$ and a set of frequencies $\omega_p$. The magnitude of the field at each frequency is represented by ${\frak E}_{p,\omega_p}$. The sum over the frequencies $\sum_\omega {\frak E}_\omega e^{-i\omega t}$, particularly in the continuous limit $\sum_\omega\to\int d\omega$, already describes any function of time. Introduction of the overall profile $\Phi(t)$ helps by dramatically reducing the number of frequencies needed to represent a given pulse profile, i.e., by describing switching the field on and off. Because each pulse is already introduced with the most general temporal profile, we will assume that pulses with different $p$ do not overlap in time. Functions $g^*_{\xi\xi'}$ are dipole matrix elements corresponding to the transitions between states to which the field couples, and 
\begin{eqnarray}\label{eq:QWG:DeltaE}
\Delta E_{\vec{\xi},\vec{\xi}'} = E_{\vec{\xi}} - E_{\vec{\xi}'} \ge 0
\end{eqnarray}
are energy gaps between the corresponding states, with the numbering convention for the states such that $\Delta E_{\vec{\xi},\vec{\xi}'}$ are positive. The driving Hamiltonian $\hat{V}(t)$ in Eq.~(\ref{eq:QWG:V}) is written in the interaction representation, i.e., in the rotating basis. This basis exactly follows the evolution of the non-driven system---that is evolution of phase for each individual state in accordance with its energy. These are build-in local rotations. In most cases qubits are formed in the rotating frame of reference to eliminate energies (and, thus, such uncontrollable rotations) from the quantum computing description as unnecessary complication.

The evolution produced by driving a quantum system with Hamiltonian (\ref{eq:QWG:V}) up to time $t_g$ can be most generally written as a time-ordered exponential integral\cite{Mahan} 
\begin{eqnarray}\label{eq:QWG:Ug}
\hat{U}_g = T e^{-i\int_0^{t_g} dt \hat{V}(t)} \equiv e^{-idt \hat{V}(t_g)}\,...\,e^{-idt \hat{V}(2dt)}e^{-idt \hat{V}(dt)}.
\end{eqnarray}
When applied to some initial state $\psi(0)$, each exponential on the right-hand side appears as a single step in evolution of a continuous time quantum walk  
\begin{eqnarray}\label{eq:QWG:Us}
\hat{U}_\text{S}(t) \to e^{-idt \hat{V}(t)}.
\end{eqnarray}
This, however, is still not very insightful---it is not generally tractable analytically and is costly to implement numerically because $\hat{V}(t)$, potentially, is a large matrix with non-trivial time-dependent coefficients and, thus, an exact diagonalization is required at each time step to compute $\hat{U}_\text{S}(t)$. It is desirable to represent $\hat{U}_\text{S}(t)$ via a finite sequence of step operators that are either time-independent or have trivial time dependence such as
\begin{eqnarray}\label{eq:QWG:Us-diag}
e^{-idt f(t)\hat{X}} = \hat{U} e^{-idt f(t)\hat{U}^\dag\hat{X}\hat{U}}\hat{U}^\dag,
\end{eqnarray}
where $\hat{U}$ is some time-independent unitary rotation and $\hat{U}^\dag\hat{X}\hat{U}$ is a diagonal matrix. Note that time-series expansion of (\ref{eq:QWG:Us}) to the second order
\begin{eqnarray}\label{eq:QWG:Us-series}
e^{-idt \hat{V}(t)} = 1 - idt \hat{V}(t) + {\cal O}(dt^2)
\end{eqnarray}
is possible and can be used as an intermediate step in analytical derivations, being exact in the limit of $dt\to 0$. However, this is not practical in numerical calculations as it will lead to very quick and dramatic loss of unitarity after just few steps, rendering the solution unphysical.

In order to simplify $\hat{U}_\text{S}(t)$, we should go back to Hamiltonian (\ref{eq:QWG:V}). Such driving Hamiltonian can be obtained from field-matter interaction Hamiltonian
\begin{eqnarray}\label{eq:QWG:V-a}
\hat{V}_p(t) = 
\frac{\Phi(t)}{\sqrt{N}}\sum_{\omega;\,i\in E\{G\}}
\left({\frak E}^*_{p,\omega_p}\hat{a}^\dag_{\omega_p} e^{i\omega_pt}+{\frak E}_{p,\omega_p}\hat{a}_{\omega_p} e^{-i\omega_pt}\right)
\left(
g^*_i \hat{c}_i
e^{-i \Delta E_it}
+
g_i \hat{c}^\dag_i
e^{i \Delta E_it}
\right)
\end{eqnarray}
in the thermodynamic, Eq.~(\ref{eq:TD-limit}), and coherent, Eqs.~(\ref{eq:QWG:a-coherent-states}), limits of bosonic field represented via creation and annihilation operators $\hat{a}^\dag_{\omega_p}$ and $\hat{a}_{\omega_p}$ for each frequency of the control field. Here we introduce rising/lowering operators 
\begin{eqnarray}\label{eq:QWG:c}
\hat{c}^\dag_i = \V{\vec{\xi}'_i}\iV{\vec{\xi}_i},
\quad\quad
\hat{c}_i = \V{\vec{\xi}_i}\iV{\vec{\xi}'_i},
\quad\quad
\hat{c}^\dag_i = [\hat{c}_i]^\dag,
\end{eqnarray}
by noticing that the summation in Eq.~(\ref{eq:QWG:V}) is in fact over the edges of the graph---each edge, $i$, of graph $G$ is represented by two vertex indices $\xi_i$ and $\xi_i'$ which it connects. Note that operators $\hat{c}_i^{(\dag)}$ with different $i$ do not commute if they share any of the vertices. When the gate aims to coherently manipulate qubits, the thermodynamic limit, Eq.~(\ref{eq:TD-limit}), is critical. The wave function of the combined system of a particle evolving on quantum network defined by graph G and bosons of the control field is given by
\begin{eqnarray}\label{eq:QWG:Psi-a}
\V{\Psi} =
\int d\alpha'^*d\alpha'\V{\alpha'}\iV
{\alpha'}
\hat{U}_\text{S}(t_g)\,...\,\hat{U}_\text{S}(2dt)\hat{U}_\text{S}(dt)
\V{\psi(0)}\V{\alpha}
\end{eqnarray}
This function is not generally factorisable, except in some special cases---bosons and the quantum particle become entangled very quickly. The entanglement becomes vanishingly small in the thermodynamic limit (\ref{eq:TD-limit})
\begin{eqnarray}\label{eq:QWG:Psi-fact}
\lim_{N\to\infty}\V{\Psi} \to \V{\psi(t_g)}\V{\alpha}.
\end{eqnarray}
Here we incorporate all acquired phases into $\V{\psi(t_g)}$. Without this limit, driving  field will introduce significant decoherence to the qubits being driven and, thus, this is the requirement that the driving field must satisfy, not an approximation. Note that development of entanglement between a quantum particle and the field in Eq.~(\ref{eq:QWG:Psi-a}) is similar to evolution of coined quantum walk described in Sec.~\ref{sec:I:QW}. This suggests that each walk in (\ref{eq:QG:Ug-psi-basis}), before the thermodynamic limit is taken, can potentially be a {\it coined} quantum walk rather than a continuous time walk. In order to investigate this further we first focus on single mode $\omega$.

\subsection{a single-mode pulse}\label{sec:QWG:1W}

The Hamiltonian (\ref{eq:QWG:V-a}) naturally splits into two parts
\begin{eqnarray}\label{eq:QWG:1W:V}
\hat{V}_\omega(t) = \hat{V}^+_\omega(t) + \hat{V}^-_\omega(t),
\end{eqnarray}
where
\begin{eqnarray}\label{eq:QWG:1W:V-}
\hat{V}^-_\omega(t) &=& 
\frac{\Phi(t)}{\sqrt{N}}\sum_{i\in E\{G\}}
\left(
{\frak E}^*_\omega
g^*_i \hat{c}_i \hat{a}^\dag
e^{-i \delta^{(-)}_it}
+
h.c.
\right),
\\\label{eq:QWG:1W:V+}
\hat{V}^+_\omega(t) &=& 
\frac{\Phi(t)}{\sqrt{N}}\sum_{i\in E\{G\}}
\left(
{\frak E}_\omega
g^*_i \hat{c}_i \hat{a}
e^{-i \delta^{(+)}_it}
+
h.c.
\right),
\end{eqnarray}
and
\begin{eqnarray}\label{eq:QWG:1W:delta}
\delta^{(\pm)}_i &=&\Delta E_i \pm \omega,
\\\label{eq:QWG:1W:delta-relation}
\delta^{(+)}_i &=& \delta^{(-)}_i + 2\omega
\end{eqnarray}
are detuning. While $\hat{V}^\pm_\omega(t)$ do not commute with each other, $\hat{U}_S(t)$ can still be split into two unitary rotations to the leading order in $dt$ using Baker-Hausdorff formula
\begin{eqnarray}\label{eq:QWG:1W:Us}
e^{dt(X+Y)} = e^{dtX}e^{dtY}e^{-dt^2[X,Y]/2}\times...
= e^{dtX}e^{dtY} +{\cal O}(dt^2).
\end{eqnarray}
We obtain
\begin{eqnarray}\label{eq:QWG:1W:Us}
\hat{U}_\text{S}(t) = e^{-idt\hat{V}^-_\omega(t)}e^{-idt\hat{V}^+_\omega(t)} + {\cal O}(dt^2).
\end{eqnarray}

Time dependence in both factors can be removed
\begin{eqnarray}\label{eq:QWG:1W:Us-H0}
\hat{U}_\text{S}(t) = e^{it\hat{H}_0} 
e^{-idt\hat{V}^-_\omega}e^{-idt\hat{V}^+_\omega}
e^{-it\hat{H}_0}
\end{eqnarray}
by introducing
\begin{eqnarray}\label{eq:QWG:1W:H0}
\hat{H}_0 = \sum_\xi E_\xi \V{\xi}\iV{\xi} + \omega \hat{a}^\dag \hat{a}.
\end{eqnarray}
The rotations due to diagonal noninteracting Hamiltonian (\ref{eq:QWG:1W:H0}) appear to depend on current time at each step (\ref{eq:QWG:1W:Us-H0}). By simple examination of several steps
\begin{eqnarray}\label{eq:QWG:1W:Us2}
\hat{U}_\text{S}(t)\hat{U}_\text{S}(t-dt) = 
e^{it\hat{H}_0} 
e^{-idt\hat{V}^-_\omega}e^{-idt\hat{V}^+_\omega}
e^{-idt\hat{H}_0} 
e^{-idt\hat{V}^-_\omega}e^{-idt\hat{V}^+_\omega}
e^{-i(t-dt)\hat{H}_0}
\end{eqnarray}
we notice that this is not the case and that they are local and do not depend on accumulated time. Re-groping the terms, we see that the gate evolution operator is given by
\begin{eqnarray}\label{eq:QWG:1W:Us2}
\hat{U}_g = \lim_{N\to\infty}e^{it_g\hat{H}_0}\hat{U}_\text{S}(t_g)...\hat{U}_\text{S}(2dt)\hat{U}_\text{S}(dt),
\end{eqnarray}
with each step defined as
\begin{eqnarray}\label{eq:QWG:1W:Us-dt}
\hat{U}_\text{S}(t) = e^{-idt\hat{V}^-_\omega}e^{-idt\hat{V}^+_\omega}e^{-idt\hat{H}_0}.
\end{eqnarray}
Note that each $\hat{V}^{\pm}_\omega$ still depend on time via slowly changing $\Phi(t)$, pulse envelop function, but this dependence is trivial, see Eq.~(\ref{eq:QWG:Us-diag}).

Equation (\ref{eq:QWG:1W:Us-dt}) defines coined quantum walk 
\begin{eqnarray}\label{eq:QWG:1W:walk}
\V{\Psi(t+dt} = \hat{S}\hat{C}\V{\Psi(t)}
\end{eqnarray}
on graph G with step operator
\begin{eqnarray}\label{eq:QWG:1W:S}
\hat{S} = e^{-idt\hat{V}^-_\omega}e^{-idt\hat{V}^+_\omega}
e^{-idt\sum_\xi E_\xi \V{\xi}\iV{\xi}},
\end{eqnarray}
and coin rotation
\begin{eqnarray}\label{eq:QWG:1W:C}
\hat{C} = e^{-idt\omega \hat{a}^\dag \hat{a}}.
\end{eqnarray}
The coin is represented by the collection of bosons participating in the control pulse and has initial state $\V{\alpha}$. The probability distribution can be formally computed as
\begin{eqnarray}\label{eq:QWG:1W:P}
P_\xi(t_f) = 
\lim_{N\to\infty} 
\iV{\xi}\iV{\alpha_N}
\hat{S}^\dag_f(t_f)\hat{C}^\dag_f(t_f)[\hat{S}\hat{C}\,...\,\hat{S}\hat{C}]
\V{\xi}\V{\alpha_N},
\end{eqnarray}
where
\begin{eqnarray}\label{eq:QWG:1W:alpha_N}
\V{\alpha_N} \overset{|\alpha|^2=N}{\equiv} \V{\alpha}
\end{eqnarray}
and
\begin{eqnarray}\label{eq:QWG:1W:S-final}
\hat{S}_f(t) &=& e^{-it\sum_\xi E \V{\xi}\iV{\xi}}
= \left[\lim_{{\frak E}\to 0}\hat{S}\right]^{t/dt},
\\\label{eq:QWG:1W:C-final}
\hat{C}_f(t) &=& e^{-it\omega \hat{a}^\dag \hat{a}} = [\hat{C}]^{t/dt}.
\end{eqnarray}

Because we are only interested in the thermodynamic limit of the above evolution we can further simplify $\hat{S}$ and $\hat{C}$. By comparing (\ref{eq:QWG:V}) and (\ref{eq:QWG:V-a}) we notice that in the thermodynamic limit $\hat{a}^\dag/\sqrt{N}$ and $\hat{a}/\sqrt{N}$ are merely placeholders for $e^{i\omega t}$ and $e^{-i\omega t}$ respectively. In other words, these operators commute
\begin{eqnarray}\label{eq:QWG:1W:a-comm}
\frac{\hat{a}}{\sqrt{N}}
\frac{\hat{a}^\dag}{\sqrt{N}}
=
\frac{\hat{a}^\dag}{\sqrt{N}}
\frac{\hat{a}}{\sqrt{N}}
+ \frac{1}{N}
\end{eqnarray}
to oder $1/N$ and, once operators $\hat{C}$ are removed, can be rearranged to push all $\hat{a}^\dag$ to the left acting on $\iV{\alpha_N}$ and all $\hat{a}$ to the right acting on $\V{\alpha_N}$, each producing $\sqrt{N}$ times a phase factor that can be lumped together with ${\frak E}_\omega$. Thus in the thermodynamic limit we can define
\begin{eqnarray}\label{eq:QWG:1W:I}
\hat{I}^{+} = \hat{a}^\dag / \sqrt{N},
\quad\quad
\hat{I}^{-} = \hat{a} / \sqrt{N},
\quad\quad
\hat{I}^+ \hat{I}^- - \hat{I}^- \hat{I}^+ = {\cal O}(1/N)\to 0,
\end{eqnarray}
and
\begin{eqnarray}\label{eq:QWG:1W:Iz}
\hat{I}^z = \hat{a}^\dag\hat{a} - N.
\end{eqnarray}

This are, in fact, operators associated with quasienergy states \cite{Zeldovich,Solenov-MOSFET,Burdov} describing the system with a periodic driving Hamiltonian, such as (\ref{eq:QWG:V}). They move quantum particle up or down the equidistant quasienergy ladder. These quasienergy states of the coin are there to accumulate the correct phase, distributing it at each step via $\hat{S}$.

\subsection{multimode pulses}\label{sec:QWG:NW}

Generalization of the above derrivation to account for multiple modes, defined by $\omega$ and ${\frak E}_\omega$, and multiple pulses, defined by $\Phi_p(t)$, is now straightforward. The summation over frequency can be restored in each term without affecting the results. Note that each frequency must have its own coin. We obtain
\begin{eqnarray}\label{eq:QWG:NW:S}
\hat{S}_p &=& e^{-idt\sum_{\omega_p} \hat{V}^-_{p,\omega_p}}e^{-idt\sum_{\omega_p} \hat{V}^+_{p,\omega_p}}
e^{-idt\sum_\xi E_\xi \V{\xi}\iV{\xi}},
\quad\quad
\hat{S}_{p,f}(t) = \left[\lim_{{\frak E}\to 0}\hat{S}_p\right]^{t/dt},
\\\label{eq:QWG:NW:C}
\hat{C}_p &=& e^{-idt\sum_{\omega_p} \omega_p 
\hat{I}^z_{\omega_p}
},
\quad\quad
\hat{C}_{p,f}(t) = [\hat{C}_p]^{t/dt},
\end{eqnarray}
with the replacement in $\hat{V}^\pm$
\begin{eqnarray}\label{eq:QWG:NW:a-cor}
\hat{a}_{\omega_p}/\sqrt{N}\to \hat{I}^-_{\omega_p},
\quad\quad
\hat{a}^\dag_{\omega_p}/\sqrt{N}\to \hat{I}^+_{\omega_p}.
\end{eqnarray}
Because we have initially introduced pulses that do not overlap in time, quantum walks evolution due to each is computed independently with the result
\begin{eqnarray}\label{eq:QWG:NW:walk}
\V{\psi(t_f)} = \iV{N}\left[ \prod_p
\lim_{N\to\infty}
\V{N}\iV{N}
\hat{S}_{p,f}^\dag(t_f)
\hat{C}^\dag_{p,f}(t_f)
\left(
\hat{S}_p\hat{C}_p...\hat{S}_p\hat{C}_p
\right)
\right]
\V{\psi(0)}\V{N},
\end{eqnarray}
where $\V{\psi(0)}$ is the initial state of the walker and
\begin{eqnarray}\label{eq:QWG:NW:N-state-omega}
\V{N} = \prod_{\omega_p}\V{\alpha_{N,\omega_p}}
\end{eqnarray}
is the initial (and final when $N\to\infty$) state of the coins.
Note that the limit must be taken after every pulse $p$, as shown in (\ref{eq:QWG:NW:walk}). Once again, we note that $N\to\infty$ limit is needed, see Eq.~(\ref{eq:QWG:Psi-fact}), to formulate quantum gates, which will be done in a later section. It is not essential to the walk itself. Note also that while $\hat{S}_{p,f}(t_f)$ does depend on the size of graph $G$, it does not depend on the composition of the pulses entering via ${\frak E}_{p,\omega_o}$, see Eq.~(\ref{eq:QWG:NW:S}).

\subsection{rotating wave approximation}\label{sec:QWG:RWA}

In many qubit architectures
\begin{eqnarray}\label{eq:QWG:1W:U+}
\omega_p \gg \Delta E_i - \omega_p
\end{eqnarray}
and the time scale defined by $1/\omega_p$ is significantly shorter (faster) than any other time scales in the problem. As the result, operators $\hat{V}^+_{p,\omega_p}$ that accumulate phase factors of $e^{-it\delta^+_{i,\omega_p}}$ rotate on much faster time scale, as compared to $\hat{V}^-_{p,\omega_p}$. Define $dt_- = n_- dt$ such that $n_-\gg 1$, but $dt_-$, defining the time scale of the slow $\hat{V}^-$ processes, is still vanishingly small, i.e., $dt_-\to 0$. We can then consider $n_-$ steps of walk (\ref{eq:QWG:NW:walk}) factoring out slow evolution as
\begin{eqnarray}\label{eq:QWG:RWA:n--steps}
\underbrace{
\hat{S}_p\hat{C}_p...\hat{S}_p\hat{C}_p
}_{n_-} =
\hat{S}^-_p\hat{C}^-_p
\left[
\hat{S}_{p,f}^{+\,\dag}(n_-dt)
\hat{C}^\dag_{p,f}(n_-dt)
\underbrace{
\hat{S}^+_p\hat{C}_p...\hat{S}^+_p\hat{C}_p
}_{n_-}
\right],
\end{eqnarray}
where
\begin{eqnarray}\label{eq:QWG:RWA:S}
\hat{S}_p^- &=& e^{-idt_-\sum_{\omega_p} \hat{V}^-_{p,\omega_p}}
e^{-idt_-\sum_\xi E_\xi \V{\xi}\iV{\xi}},
\quad\quad
\hat{S}^-_{p,f}(t) = \left[\lim_{{\frak E}\to 0}\hat{S}^-_p\right]^{t/dt},
\\\label{eq:QWG:RWA:C}
\hat{C}^-_p &=& e^{-idt_-\sum_{\omega_p} \omega_p 
\hat{I}^z_{\omega_p}
},
\quad\quad
\hat{C}^-_{p,f}(t) = [\hat{C}^-_p]^{t/dt},
\end{eqnarray}
and
\begin{eqnarray}\label{eq:QWG:RWA:U+}
\hat{S}^+_p = 
e^{-idt\sum_{\omega_p} \hat{V}^+_{p,\omega_p}}
e^{-idt\sum_\xi E_\xi \V{\xi}\iV{\xi}},
\quad\quad
\hat{S}^+_{p,f}(t) = \left[\lim_{{\frak E}\to 0}\hat{S}^+_p\right]^{t/dt}.
\end{eqnarray}
For large enough $n_-$ rotations
\begin{eqnarray}\label{eq:QWG:RWA:n--steps}
\hat{S}_{p,f}^{+\,\dag}(n_-dt)
\hat{C}^\dag_{p,f}(n_-dt)
\underbrace{
\hat{S}^+_p\hat{C}_p...\hat{S}^+_p\hat{C}_p
}_{n_-}
\to 1,
\end{eqnarray}
i.e., this auxiliary walk effectively averages itself to 1 and we obtain
\begin{eqnarray}\label{eq:QWG:RWA:walk}
\V{\psi(t_f)} = \iV{N}\left[ \prod_p
\lim_{N\to\infty}
\V{N}\iV{N}
\hat{S}_{p,f}^{-\dag}(t_f)
\hat{C}^{-\dag}_{p,f}(t_f)
\left(
\hat{S}^-_p\hat{C}^-_p...\hat{S}^-_p\hat{C}^-_p
\right)
\right]
\V{\psi(0)}\V{N}.
\end{eqnarray}

\subsection{resonant approximation}\label{sec:QWG:RA}

In many systems where rotating wave approximation is applicable, the time scale defined by $dt_-$ can be further split into resonant and non-resonant time scales. The latter is still defined by $dt_-$ and is due to the set of non-zero detunings $\{\delta^-_{i,\omega_p}\neq 0\}$. A much slower time scale is associated with the subset of $\{{\frak E}_{p,\omega_p}g_i\}$ for which $\delta^-_{i,\omega_p}=0$. Introducing $dt_R = n_R dt_-$ to follow the slow scale, such that $n_R\gg 1$, but $dt_R\to 0$, we can rearrange $n_R$ steps of walk (\ref{eq:QWG:RWA:walk}) as 
\begin{eqnarray}\label{eq:QWG:RA:nR-steps}
\underbrace{
\hat{S}^-_p\hat{C}^-_p...\hat{S}^-_p\hat{C}^-_p
}_{n_-} =
\hat{S}^{\in R}_p\hat{C}^R_p
\left[
\hat{S}^{-\,\not\in R}_{p,f}(n_Rdt_-)^\dag
\hat{C}^{-\dag}_{p,f}(n_Rdt_-)
\underbrace{
\hat{S}^{-\,\not\in R}_p\hat{C}^-_p...\hat{S}^{-\,\not\in R}_p\hat{C}^-_p
}_{n_R}
\right],
\end{eqnarray}
where
\begin{eqnarray}\label{eq:QWG:RA:S}
\hat{S}_p^{\in R} &=& e^{-idt_R\sum_{\omega_p} \hat{V}^{-,\in R}_{p,\omega_p}}
e^{-idt_R\sum_\xi E_\xi \V{\xi}\iV{\xi}},
\quad\quad
\hat{S}^{\in R}_{p,f}(t) = \left[\lim_{{\frak E}\to 0}\hat{S}^{\in R}_p\right]^{t/dt}
\\\label{eq:QWG:RA:C}
\hat{C}^R_p &=& e^{-idt_R\sum_{\omega_p} \omega_p 
\hat{I}^z_{\omega_p}
},
\quad\quad
\hat{C}^R_{p,f}(t) = [\hat{C}^R_p]^{t/dt},
\end{eqnarray}
and
\begin{eqnarray}\label{eq:QWG:RA:UnotR}
\hat{S}_p^{-\,\not\in R} = e^{-idt_-\sum_{\omega_p} \hat{V}^{-,\not\in R}_{p,\omega_p}}
e^{-idt_-\sum_\xi E_\xi \V{\xi}\iV{\xi}},
\quad\quad
\hat{S}^{-\,\not\in R}_{p,f}(t) = \left[\lim_{{\frak E}\to 0}\hat{S}^{-\,\not\in R}_p\right]^{t/dt}.
\end{eqnarray}
Here the sum over $i$ and $\omega_p$ in matrix $\hat{V}^{-}_{p,\omega_p}$ was split into two terms
\begin{eqnarray}\label{eq:QWG:RA:V}
\hat{V}^{-}_{p,\omega_p} = \hat{V}^{-,\in R}_{p,\omega_p} +  \hat{V}^{-,\not\in R}_{p,\omega_p},
\end{eqnarray}
depending on whether a given term $i$ has $\Delta E_i$, corresponding to the $i$-th edge of graph G, equal to $\omega_p$, in which case the term is labeled as ``$\in R$'' term, or not, in which case it is labeled as ``$\not\in R$''. The exponential of $\hat{V}^{-}_{p,\omega_p}$ can then be factored out into two to order ${\cal O}(dt_R^2)$. By noting that all terms in the left-hand side of (\ref{eq:QWG:RA:nR-steps}) commute to order ${\cal O}(dt_R^2)$ we obtain Eq.~(\ref{eq:QWG:RA:nR-steps}).

As earlier, $n_R$-step non-resonant walk averages to 1
\begin{eqnarray}\label{eq:QWG:RA:not-inR-steps}
\hat{S}^{-\,\not\in R}_{p,f}(n_Rdt_-)^\dag
\hat{C}^{-\dag}_{p,f}(n_Rdt_-)
\underbrace{
\hat{S}^{-\,\not\in R}_p\hat{C}^-_p...\hat{S}^{-\,\not\in R}_p\hat{C}^-_p
}_{n_R}
\to 1.
\end{eqnarray}
This effectively resets the coins on every step, $dt_R$, of the slow time scale, which can be thought of as Markovian approximation of removing memory effects in the environment (coins). The averaging depends on separation of scales of resonant and non-resonant evolution. It will not occur if such separation is insufficient, resulting in decoherence if coins' degrees of freedom are subsequently traced out. If separation of scale is sufficient to guarantee (\ref{eq:QWG:RA:not-inR-steps}) to desired accuracy we obtain
\begin{eqnarray}\label{eq:QWG:RA:walk-complex}
\V{\psi(t_f)} = \iV{N}\left[ \prod_p
\lim_{N\to\infty}
\V{N}\iV{N}
\hat{S}_{p,f}^{\in R \dag}(t_f)
\hat{C}^{\in R \dag}_{p,f}(t_f)
\left(
\hat{S}^{\in R}_p\hat{C}^R_p...\hat{S}^{\in R}_p\hat{C}^R_p
\right)
\right]
\V{\psi(0)}\V{N}
\end{eqnarray}

Quantum walk (\ref{eq:QWG:RA:walk-complex}) can be simplified further by noticing that rotation $e^{-idt_R\sum_\xi E_\xi \V{\xi}\iV{\xi}}$ counters all the phase introduced to the walker by $\hat{C}_p^R$ in all steps, including finally with the product $\hat{S}_{p,f}^{\in R \dag}(t_f)\hat{C}^{\in R \dag}_{p,f}(t_f)$. This is a manifestation of the fact that all entries in matrix $\hat{V}^{-,\in R}_{p,\omega_p}$ were chosen such that $\Delta E_i$ is equal to one of the frequencies $\omega_p$. Therefore operators $\hat{I}^\pm_{\omega_p}$ located there introduce no phase. They simply shift the appropriate coin states and can be removed without any change in the thermodynamic limit of $N\to\infty$. As the result, evolutions of the coin and the walker are completely independent---the coin is factored out and can be removed. The remaining $\hat{S}^{\in R}_p$ rotations commute and can be re-combined as follows
\begin{eqnarray}\label{eq:QWG:RA:S-tau}
\hat{S}^{\in R}_p...\hat{S}^{\in R}_p = 
e^{-i\int\limits_0^{t_g} dt\Phi(t)\hat{\Lambda}}
= e^{-i\tau_g\hat{\Lambda}} 
= \left[e^{-id\tau\hat{\Lambda}}\right]^{[t_g/d\tau]},
\end{eqnarray}
with the redefinition of a single step to
\begin{eqnarray}\label{eq:QWG:RA:Us}
\hat{U}^{\in R}_S = e^{-id\tau\hat{\Lambda}}.
\end{eqnarray}
Here
\begin{eqnarray}\label{eq:QWG:RA:tau}
\tau = \int\limits_0^{t} dt\Phi(t),
\quad\quad
d\tau = dt\Phi(t),
\end{eqnarray}
where the overall pulse profile $\Phi(t)$ enters as a metric connecting real and effective time axes.
The walk is given by
\begin{eqnarray}\label{eq:QWG:RA:walk}
\V{\psi(t_f)} = 
[\hat{U}_\text{S}^{\in R}]^{[t_g/d\tau]} \V{\psi(0)}
=
e^{-i\tau\hat{\Lambda}}\V{\psi(0)}
\end{eqnarray}
This is a continuous time quantum walk with time-independent adjacency matrix $\Lambda$ propagating in time $\tau$.

\section{Quantum walk on a line}\label{sec:QWG-1D}

Solving coined or continuous time quantum walks on general large graphs is difficult. Analytical solutions can be obtained in some special cases, such as the case of one-dimensional (1D) graphs that are chains of states of different lengths. For example, without any symmetry on edges, continuous time quantum walks on 1D chains can be solved analytically and explicitly for chains of up to 5 states~---~4 complex parameters (one for each edge) forming a general polynomial equation of degree 4. Longer chains can often be reduced to solvable polynomials if the system is sufficiently symmetric.

Larger graphs and respective quantum walks can be reduced to 1D chains by performing rotations of basis in appropriate parts of the graph.\cite{Householder,Cavin-Solenov,Novo} In order to benefit from this approach when constructing quantum gates we must make sure that initial and final sets of nodes (basis states) still reside in qubit domain, i.e. on graph $G_Q$. They should remain unchanged, or, at least, do not mix with any of the intermediate nodes residing in $G-G_Q$. Such transformation can be formally written as
\begin{eqnarray}\label{eq:QWG-1D:L}
&&{\cal L}^T_{E(G-G_Q)}\{ \hat{X} \} \equiv \hat{T}\hat{X} \hat{T}^\dag.
\end{eqnarray}
It produces a superposition for each edge transition operator
\begin{eqnarray}\label{eq:QWG-1D:Lc}
\sum_{i}g^*_i
{\cal L}^T_{E(G-G_Q)}\{ \hat{c}_i \}
=\sum_{i,i'}g^*_iT_{\eta_{i'},\xi_i}T^*_{\eta'_{i'},\xi'_i}
\V{\eta_{i'}}\iV{\eta'_{i'}}
=\sum_{i'}
\left[
\sum_{i}g^*_iT_{\eta_{i'},\xi_i}T^*_{\eta'_{i'},\xi'_i}
\right]
\hat{c}'_{i'}.
\end{eqnarray}
Multiple simple transformations can be done consecutively to reduce the graph. Examples are given in Sec.~\ref{sec:Gs} and Appendix~\ref{sec:SG}.

In all discussed walks, i.e., Eqs.~(\ref{eq:QWG:NW:walk}), (\ref{eq:QWG:RWA:walk}), and (\ref{eq:QWG:RA:walk}), such basis rotation directly affects only $\hat{S}_p$. It enters via
\begin{eqnarray}\label{eq:QWG-1D:V+-}
e^{-idt\sum_{\omega_p} \hat{V}^\pm_{p,\omega_p}}
\to
e^{-idt\sum_{\omega_p} {\cal L}^T_{E(G-G_Q)}\{
\hat{V}^\pm_{p,\omega_p}
\}
}
=
e^{-idt\sum_{i'\omega_p} \left[
{\frak E}^{(*)}_{p,\omega_p}g'^*_{i'}
\hat{c}_{i'}
\hat{I}^\pm_{\omega_p}
+h.c.
\right]
},
\end{eqnarray}
\begin{eqnarray}\label{eq:QWG-1D:g'i'}
g'^*_{i'} =
\sum_{i} g^*_i
T_{\eta_{i'},\xi_i}T^*_{\eta'_{i'},\xi'_i},
\end{eqnarray}
In the case of Eqs.~(\ref{eq:QWG:NW:walk}) and (\ref{eq:QWG:RWA:walk}), it also affects the non-coined phase term in $\hat{S}_p$
\begin{eqnarray}\label{eq:QWG-1D:H0}
e^{-idt
\sum_{\xi} E_\xi \V{\xi}\iV{\xi}
}
\to
e^{-idt
\sum_{\xi} E_\xi 
\hat{T}\V{\xi}\iV{\xi}\hat{T}^\dag
}
=
e^{-idt
\sum_{\xi,\eta,\eta'} E_\xi 
T_{\eta,\xi}T^*_{\eta',\xi'}
\V{\eta}\iV{\eta'}
}
\end{eqnarray}
making it non-diagonal. Thus, in the case of walks (\ref{eq:QWG:NW:walk}) and (\ref{eq:QWG:RWA:walk}), such rotations potentially add additional non-trivial continuous time walk rotation at every step. The graph for the coined walk is $G'$ as obtained after transformation ${\cal L}^T_{E(G-G_Q)}$. The additional non-coined continuous time quantum walk evolves on graph $G''$ obtained from the right-hand side of Eq.~(\ref{eq:QWG-1D:H0}).  It is not equal to $G$ or $G'$.
Thus, ${\cal L}^T_{E(G-G_Q)}$ must therefore be chosen to balance the complexity of $G'$ and $G''$ to simplify the overall solution. Fortunately, as we will see later, in many cases $G''$ is still a simple disjointed graph with only few 1D segments and all other vertices unconnected (except for energy self-loops).
The phase term (\ref{eq:QWG-1D:H0}) is not present for resonant walk and the transformation is applied to $\hat\Lambda$ in Eq.~(\ref{eq:QWG:RA:Us}) simplifying the solution directly without any side effects.

\section{More general field-matter interactions}\label{sec:QWG-GEN}

In some cases, e.g., for trapped ions qubit architectures,\cite{Blinov,CiracZoller} a different interaction with bosonic control field is necessary. In order to show that the quantum walk description obtained above is still applicable, we generalize interaction (\ref{eq:QWG:V-a}) to
\begin{eqnarray}\label{eq:QWG-GEN:V-a}
\hat{V}_p = 
\Phi(t)\sum_{\omega;\,i\in E\{G\}}
{\frak F}
\left(
{\frak E}^*_{p,\omega_p}\frac{\hat{a}^\dag_\omega}{\sqrt{N}}
+
{\frak E}_{p,\omega_p}\frac{\hat{a}_\omega}{\sqrt{N}} 
\right)
\left(
g^*_i \hat{c}_i
+
g_i \hat{c}^\dag_i
\right),
\end{eqnarray}
where
\begin{eqnarray}\label{eq:QWG-GEN:f(x)}
{\frak F}(x) = \sum_{m=0}^\infty
\frac{{\frak F}^{(m)}(0)}{m!} x^m.
\end{eqnarray}
is an arbitrary Taylor-series expandable function. Using the replacement suggested earlier in Eqs.~(\ref{eq:QWG:1W:I}) and (\ref{eq:QWG:NW:a-cor}) for the thermodynamic limit we obtain
\begin{eqnarray}\label{eq:QWG-GEN:V-I}
\hat{V}_p = 
\Phi(t)\sum_{m,\omega;\,i\in E\{G\}}
\frac{{\frak F}^{(m)}(0)}{m!}
\left(
{\frak E}^*_{p,\omega_p}\hat{I}^+_\omega
+
{\frak E}_{p,\omega_p}\hat{I}^-_\omega 
\right)^m
\left(
g^*_i \hat{c}_i
+
g_i \hat{c}^\dag_i
\right).
\end{eqnarray}
Binomial expansion and relations in Eq.~(\ref{eq:QWG:1W:I}) can be used to split rising and lowering operators into two terms. We obtain
\begin{eqnarray}\label{eq:QWG-GEN:x-m}
&&\left(
{\frak E}^*_{p,\omega_p}\hat{I}^+_\omega
+
{\frak E}_{p,\omega_p}\hat{I}^-_\omega
\right)^m
=
\sum_{k=0}^m {m\choose k}
{\frak E}^{*\,k}_{p,\omega_p} {\frak E}^{m-k}_{p,\omega_p}
[\hat{I}^+_\omega]^k [\hat{I}^-_\omega]^{m-k}
\\\nonumber
&&=
\sum_{k\le m/2} {m\choose k}
{\frak E}^{*\,k}_{p,\omega_p} {\frak E}^{m-k}_{p,\omega_p}
[\hat{I}^-_\omega]^{m-2k}
+
\sum_{k>m/2} {m\choose k}
{\frak E}^{*\,k}_{p,\omega_p} {\frak E}^{m-k}_{p,\omega_p}
[\hat{I}^+_\omega]^{2k-m}
\\\nonumber
&&=
\delta_{m\in even}{m\choose m/2}
|{\frak E}_{p,\omega_p}|^m
+
\left(
\sum_{k>m/2} 
{m\choose k}
{\frak E}^{*\,k}_{p,\omega_p} {\frak E}^{m-k}_{p,\omega_p}
[\hat{I}^+_\omega]^{2k-m}
+h.c.
\right)
\end{eqnarray}

The derived quantum walk description, see Eqs.~(\ref{eq:QWG:NW:walk}), (\ref{eq:QWG:RWA:walk}), and (\ref{eq:QWG:RA:walk}), is based on $\hat{V}^\pm_p$ defined in Eqs.~(\ref{eq:QWG:1W:V-}) and (\ref{eq:QWG:1W:V+}). For a general form of coupling to control field they must be redefined to
\begin{eqnarray}\label{eq:QWG-GEN:V-}
&&\hat{V}^-_p 
= 
\Phi(t)\sum_{\omega;\,i\in E\{G\}}
\sum_{m\in even,k>m/2}
\frac{{\frak F}^{(m)}(0)}{m!}
{m\choose k}|{\frak E}_{p,\omega_p}|^m
\left(
g^*_i \hat{c}_i
+
h.c.
\right)
\\\nonumber
&&+
\Phi(t)\sum_{\omega;\,i\in E\{G\}}
\sum_{m,k>m/2}
\left(
\frac{{\frak F}^{(m)}(0)}{m!}
{m\choose k}
{\frak E}^{*\,k}_{p,\omega_p} {\frak E}^{m-k}_{p,\omega_p}
g^*_i \hat{c}_i
[\hat{I}^+_\omega]^{2k-m}
+
h.c.
\right)
\end{eqnarray}
and
\begin{eqnarray}\label{eq:QWG-GEN:V+}
\hat{V}^+_p 
=
\Phi(t)\sum_{\omega;\,i\in E\{G\}}
\sum_{m,k>m/2}
\left(
\frac{{\frak F}^{(m)}(0)}{m!}
{m\choose k}
{\frak E}^{k}_{p,\omega_p} {\frak E}^{*\,m-k}_{p,\omega_p}
g^*_i \hat{c}_i
[\hat{I}^-_\omega]^{2k-m}
+
h.c.
\right)
\end{eqnarray}
All other derivations remain the same. While Eq.~(\ref{eq:QWG-GEN:V-}) has a sum over powers of $\hat{I}^+$ and $\hat{I}^-$ not present in Eq.~(\ref{eq:QWG:1W:V-}) beyond linear terms, it is still straightforward to split it into resonant and non-resonant parts.

\section{Gates by quantum walks}\label{sec:Gs}

We begin here by summarizing the above derivations.
As outlined in Eq.~(\ref{eq:QG:Ug-psi-basis}), quantum gates are rotations of the basis states of a single or multi-qubit sub-system. These rotations can be represented by quantum walks derived in the previous sections as
\begin{eqnarray}\label{eq:Gs:walk}
&&\V{\xi_1\xi_2...}'_{\in G_Q}
= 
\left[ \prod_p
\lim_{N\to\infty}
\iV{N}
S_{p,f}^\dag
C^\dag_{p,f}
[
\underbrace{
\hat{S}_p\hat{C}_p\,...\,\hat{S}_p\hat{C}_p\hat{S}_p\hat{C}_p
}_{t_g/dt}
]
\V{N}
\right]
\V{\xi_1\xi_2...}_{\in G_Q},
\\\label{eq:Gs:gate}
&&
\V{\xi_1\xi_2...}'_{\in G_Q}
= \hat{U}_g \V{\xi_1\xi_2...} 
\equiv
\sum_{\xi_1'\xi_2'...\in G_Q}
U_{\xi_1'\xi_2'...,\xi_1\xi_2...}\V{\xi_1'\xi_2'...},
\end{eqnarray}
where $U_{\xi_1'\xi_2'...,\xi_1\xi_2...}$ are entries in the $\hat{U}_g$ matrix in a given basis. Each quantum walk begins from one of the qubit basis states $\V{\xi_1\xi_2...}$, propagates (in general) through the entire available quantum network and returns to state $\V{\xi_1\xi_2...}'$ in $G_Q$ that is a specific superposition of qubit states as defined by the gate in Eq.~(\ref{eq:Gs:gate}). All walks must return completely and at the same time. When a single pulse is used to achieve that, we have a set of walks $W_g\in U_g$, each defined as
\begin{eqnarray}\label{eq:Gs:walk-ends}
\lim_{N\to\infty}
\iV{N}
\iV{\xi_1\xi_2...}'_{\in G_Q}
\hat{S}_{p,f}^\dag
\hat{C}^\dag_{p,f}
[
\underbrace{
\hat{S}_p\hat{C}_p\,...\,\hat{S}_p\hat{C}_p\hat{S}_p\hat{C}_p
}_{t_g/dt}
]
\V{N}
\V{\xi_1\xi_2...}_{\in G_Q}
=1,
\end{eqnarray}
where
\begin{eqnarray}\label{eq:Gs:N}
\V{N} \equiv \prod_{\omega_p}\V{\alpha_{N,\omega_p}}
\end{eqnarray}
is the combined state of all quantum coins realizing the walk,  and
\begin{eqnarray}\label{eq:Gs:N}
\hat{S}_{p,f} = \left[\lim_{\frak E_{p,\omega_p}\to 0}\hat{S}_{p}\right]^{t_g/dt},
\quad\quad
\hat{C}_{p,f} = \left[\hat{C}_{p}\right]^{t_g/dt}
\end{eqnarray}
are control-field-independent final phase adjustments. One quasienergy (or bosonic) coin per each control field frequency is required. The coin operator is defined as
\begin{eqnarray}\label{eq:Gs:C}
\hat{C}_p = 
e^{-idt\sum\limits_{\omega_p} \omega_p 
\hat{I}^z_{\omega_p}
},
\end{eqnarray}
as explained in Section~\ref{sec:QWG}. When no approximations are used, the walks advance via shift operator
\begin{eqnarray}\label{eq:Gs:S}
\hat{S}_p = 
e^{-id\tau\hat{\Lambda}}\,
e^{-id\tau\!\!\!\!\!\!\!\sum\limits_{\omega_p\not\in R_i,i\in E\{G\}}\!\!
\left(
{\frak E}^*_{\omega_p}
g^*_i \hat{c}_i \hat{I}^+_{\omega_p}
+
h.c.
\right)
}
e^{-id\tau\!\!\!\!\!\sum\limits_{\omega_p,i\in E\{G\}}\!\!
\left(
{\frak E}_{\omega_p}
g^*_i \hat{c}_i \hat{I}^-_{\omega_p}
+
h.c.
\right)
}
e^{-idt\sum\limits_\xi E_\xi \V{\xi}\iV{\xi}},
\end{eqnarray}
where $R_i$ denotes frequencies that are (exactly) in resonance with the transition at edge $i$, and
\begin{eqnarray}\label{eq:Gs:c}
\hat{c}_i = \V{\xi_i}\iV{\xi'_i},
\quad\quad
\hat{c}^\dag_i = \V{\xi'_i}\iV{\xi_i},
\quad\quad
\xi_i,\xi'_i \in [E\{G\}]_i,
\quad\quad
\V{\xi} \in G
\end{eqnarray}
are jump operators associated with each ($i$-th) edge of graph G, as mentioned earlier. Note that, except for the last (diagonal) phase factor, the shift operator propagates in its own time $\tau$ defined by 
\begin{eqnarray}\label{eq:Gs:tau}
d\tau = \Phi(t)dt,
\end{eqnarray}
with the metric given by the overall pulse envelop profile. 

When {\bf rotating wave approximation} is appropriate, the shift operator simplifies to
\begin{eqnarray}\label{eq:Gs:S-RWA}
\hat{S}_p \to 
e^{-id\tau\hat{\Lambda}}\,
e^{-id\tau\!\!\sum\limits_{\omega_p\not\in R_i,i\in E\{G\}}\!\!
\left(
{\frak E}^*_{\omega_p}
g^*_i \hat{c}_i \hat{I}^+_{\omega_p}
+
h.c.
\right)
}
e^{-idt\sum\limits_\xi E_\xi \V{\xi}\iV{\xi}}.
\end{eqnarray}
In this case coined non-counter-rotating term, the third exponential in Eq.~(\ref{eq:Gs:S}), rotates too quickly providing a negligible contribution to the final result, as can be estimated by quantum walk (\ref{eq:QWG:RWA:n--steps}) as explained in Sec.~\ref{sec:QWG}. Physically this happens when control field frequencies are much larger then frequencies associated with the field amplitudes (Rabi frequencies), i.e., $\omega_p \gg {\frak E}_{p,\omega_p}g_i$ and $\omega_p \gg d\Phi(t)/dt$. In most optically controlled qubit systems they differ by many orders of magnitude and, thus, this approximation is nearly exact.

In {\bf resonant approximation} only frequencies that are in exact resonance with some transitions, $\omega_p\in R_i$, are included. In this case all coins evolve independently and can be factored out. The shift operator simplifies to bare minimum
\begin{eqnarray}\label{eq:Gs:S-RA}
\hat{S}_p \to 
e^{-id\tau\hat{\Lambda}},
\quad\quad\quad
\hat{\Lambda} =
\sum\limits_{i\in E\{G\}}\!\!
\left(
{\frak E}^*_{\omega_p\in R_i}
g^*_i \hat{c}_i
+
h.c.
\right),
\end{eqnarray}
and Eq.~(\ref{eq:Gs:walk}) is simply a collection of continuous time quantum walks
\begin{eqnarray}\label{eq:Gs:walk-RA}
\V{\xi_1\xi_2...}'_{\in G_Q}
= 
e^{-id\tau\Lambda}
\V{\xi_1\xi_2...}_{\in G_Q},
\end{eqnarray}
chosen to satisfy the gate. This approximation is valid when all non-zero detunings $\delta^-_i$, defined in Eq.~(\ref{eq:QWG:1W:delta}) are much greater than frequencies associated with control field amplitudes, i.e., $\delta^-_i \gg {\frak E}_{p,\omega_p}g_i'$ and $\delta^-_i \gg d\Phi(t)/dt$. This can be verified by performing walk (\ref{eq:QWG:RA:nR-steps}). It is applicable in many cases, but the inequalities must be carefully verified for all transitions that can be potentially affected by the control field. Thus, the validity of the approximation depends heavily on the details of specific qubit architecture.

In all cases, with or without approximations, quantum walks (\ref{eq:Gs:walk}) performing the gate are governed primarily by resonant,
\begin{eqnarray}\label{eq:Gs:Omega_i}
\Omega_i\equiv g_i{\frak E}_{p,\omega_p\in R_i},
\end{eqnarray}
and non-resonant,
\begin{eqnarray}\label{eq:Gs:non-resonant-Omega}
g_i{\frak E}_{p,\omega_p\not\in R_i}
\end{eqnarray}
amplitudes of the control field at each edge of graph $G$. The former define $\hat{\Lambda}$ in each $\hat{S}_p$.
Because control field is shaped by the user, these amplitudes are adjustable complex parameters. Note that $g_i$ are not the property of the control field and, thus, are not in general adjustable (although, they can be tunable in some cases via other control mechanisms, e.g., by changing confinement that defines basis states in $G$). The parameters are chosen to ensure that Eq.~(\ref{eq:Gs:walk-ends}) is satisfied for all qubit basis sates on which the given gate is defined. This {\it is not always possible}. It depends on physical interactions between different parts of quantum network described by graph $G$. When graph $G$ is simply a collection of non-interacting qubits (even with additional states per each qubit), Eq.~(\ref{eq:Gs:walk-ends}) based on any entangling gate can not be satisfied irrespective of control field used. Mathematically, physical interactions in graph $G$ enter via symmetries in the adjacency matrix $\hat{\Lambda}$. The symmetries are defined as 
\begin{eqnarray}\label{eq:Gs:symmetry}
\Omega_i = {\frak s}_{i_0}^{\forall j\in {\cal S}_{i_0}}\Omega_i,
\end{eqnarray}
where ${\frak s}_{i_0}^j$ is the symmetry operator connecting amplitudes $\Omega_i$ at different graph edges
\begin{eqnarray}\label{eq:Gs:symmetry}
{\frak s}_{i_0}^j\Omega_{i_0} = \Omega_{i_0+j}\quad \forall\omega_p,
\quad\quad
\{{\frak s}_{i_0}^{j_1},{\frak s}_{i_0}^{j_2},...\} = {\cal S}_{i_0}.
\end{eqnarray}
All such operators form group ${\cal S}_{i_0}$ that defines a set of identical edges with edge $i_0$ as one of the elements. 
Note that this symmetry merely distributes amplitudes between resonant (\ref{eq:Gs:Omega_i}) and non-resonant (\ref{eq:Gs:non-resonant-Omega}) sets. Because evolution produced by each set is very different, such distribution directly affects all quantum walks (\ref{eq:Gs:walk}) whether or not rotating wave or resonant approximations are made. This is emphasized by the first factor in Eqs.~(\ref{eq:Gs:S}), (\ref{eq:Gs:S-RWA}), and (\ref{eq:Gs:S-RA}). The symmetries ${\cal S}_{i}$ are the greatest for a non-interacting system that have no physical interactions between qubits (or extended qubit systems)
\begin{eqnarray}\label{eq:Gs:S=S0}
{\cal S}_{i} = {\cal S}^Q.
\end{eqnarray}
In this case each index in $\V{\xi\xi'
...}$ simply numbers the sate in each qubit system (note that $\xi$ can be lager than 1 when auxiliary states are available). When physical interactions are present, we can still mark all basis states by non-interacting indexes, e.g., $\V{\xi\xi'...}=\V{01...}$, for convenience, assuming adiabatic connection with non-interacting case. However symmetries ${\cal S}_{i}$ will change because $g_i$ and $\Delta E_i$ are different. The graph (edges) will be less symmetric
\begin{eqnarray}\label{eq:Gs:S<S0}
{\cal S}_{i} < {\cal S}^Q
\end{eqnarray}
This symmetry reduction is discussed in details in Ref.~\onlinecite{Solenov-QW} within resonant approximation, i.e., as it applies to adjacency matrix $\hat{\Lambda}$. Here we see that it is a general requirement for entangling gates irrespective of approximations made. In what follows we give few examples of single and two-qubit gates with all walks optimized analytically within resonant approximation. Some other examples based on resonant approximation can be found in Ref.~\onlinecite{Solenov-QW}.

\subsection{single-qubit gates: Z, Hadamard}\label{sec:Gs:1QB}

Single-qubit Z gate has the simplest quantum walk description. When resonant approximation is appropriate, connecting one auxiliary state with one of the qubit states is sufficient. In this case adjacency matrix written in the basis $\{\V{2},\V{1},\V{0}\}$ is
\begin{eqnarray}\label{eq:Gs:1QB:Lambda}
\hat{\Lambda} = 
\MxRRR{0 & 0 & \Omega}
      {0 & 0 & 0}
      {\Omega^* & 0 & 0}
\end{eqnarray}
The single complex parameter $\Omega$ must be set to organize a non-trivial return walk
\begin{eqnarray}\label{eq:Gs:1QB:walk-Z}
\text{walk}:
\V{0} \xrightarrow[]{{\cal R}_\pi\text{ via }\V{2}} 
\V{0}
\quad\quad
e^{-i\tau\hat{\Lambda}}\V{0} = \V{0}
\end{eqnarray}
at time $\tau_g$. This is accomplished if $\tau_g\Omega = \pi (2n+1)$, with $n\in \mathbb{Z}$, as follows from evaluating the exponential of $i\tau_g\Lambda$ for this effectively two-state system. The resulting gate shown in the basis $\{\V{1},\V{0}\}$ is
\begin{eqnarray}\label{eq:Gs:1QB:Z}
\hat{U}_g({\rm Z}) \equiv \hat{\rm Z} = 
\MxQ{1 & 0}
    {0 &-1}
\end{eqnarray}
In this example we see that quantum walks simply describe the standard single-qubit control via a single leg of a ``$\Lambda$'' system (a three-state quantum system). 

Another example relying on the same ``$\Lambda$'' system, the Hadamard gate, can be realized if two adjustable amplitudes (corresponding to two resonant frequencies) are present in the control pulse. In the basis $\{\V{1},\V{2},\V{0}\}$ we have
\begin{eqnarray}\label{eq:Gs:1QB:Lambda}
\hat{\Lambda} = 
\MxRRR{0 & \Omega_1 & 0}
      {\Omega_1^* & 0 & \Omega_2^*}
      {0 & \Omega_2 & 0}
\end{eqnarray}
This time we need two continuous time quantum walks to occur (and terminate) at the same (effective) time
\begin{eqnarray}\label{eq:Gs:1QB:walk-H}
&\text{walk 1}:& 
\V{0} \xrightarrow[]{\text{via }\V{2}} 
\frac{\V{0}+\V{1}}{\sqrt{2}}
\quad\quad
e^{-i\tau\Lambda}\V{0} = \frac{\V{0}+\V{1}}{\sqrt{2}}
\\
&\text{walk 2}:& 
\V{1} \xrightarrow[]{\text{via }\V{2}} 
\frac{\V{0}-\V{1}}{\sqrt{2}}
\quad\quad
e^{-i\tau\Lambda}\V{1} = \frac{\V{0}-\V{1}}{\sqrt{2}}
\end{eqnarray}
This is accomplished by setting $|\Omega_1|/|\Omega_2| = \sqrt{2}-1$, ${\rm arg}\Omega_1 - {\rm arg}\Omega_2 = \pi$, and $\tau\sqrt{|\Omega_1|^2+|\Omega_2|^2} = \pi(2n+1)$, with $n\in\mathbb{Z}$. This result can be obtained from the exact solution, see, e.g., Sec.~5.2 in Ref.~\onlinecite{Solenov-QW}, and can be verified by direct exponentiation. The resulting gate operator is
\begin{eqnarray}\label{eq:Gs:1QB:H}
\hat{U}_g({\rm H}) \equiv \hat{\rm H} = 
\frac{1}{\sqrt{2}}
\MxQ{1 & 1}
    {1 & -1}
\end{eqnarray}

When single-qubit gates are performed in a multiqubit register, all edges involved in the gates must either have the highest symmetry (${\cal S}_i={\cal S}^0$), i.e., must not be part of the network affected by interactions, or be part of walks designed such that the reduced symmetry does not affect the propagation. Note that symmetry ${\cal S}_i$ can always be made higher artificially by choosing appropriate values for pulse amplitudes, but it can not be made lower. In contrast, multiqubit entangling gates performed on the same network must engage edges that are affected by interactions and, thus, have lower symmetry, i.e., ${\cal S}_i<{\cal S}^0$. In this case walks must be designed to probe this symmetry reduction as shown in the next subsection.

\subsection{two-qubit gates: Control-Z}\label{sec:Gs:2QB}

\begin{figure}\begin{center}
\includegraphics[width=0.4\textwidth]{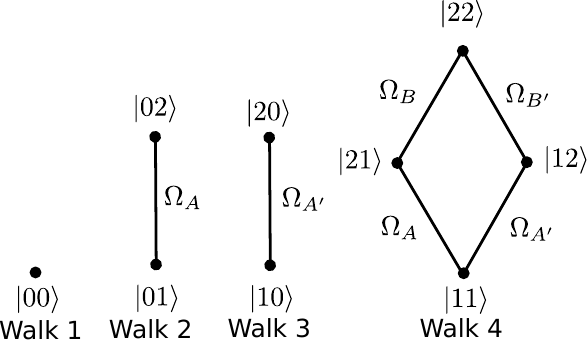}
\caption{\label{fig:CZ}
Quantum walks involved in CZ gate.
}
\end{center}\end{figure}

The simplest example of an entanglement-manipulating two-qubit gate is a CZ gate 
\begin{eqnarray}\label{eq:Gs:2QB:CZ}
\hat{U}_g({\rm CZ}) \equiv \hat{\rm CZ} = 
\MxQQ{1 & 0 & 0 & 0}
     {0 & 1 & 0 & 0}
     {0 & 0 & 1 & 0}
     {0 & 0 & 0 &-1}.
\end{eqnarray}
When qubits are in direct vicinity, it can be performed by walks propagating in subgraph G defined on states $\V{\xi\xi'}$ with $\xi,\xi'=0,1,2$. Here we assume that the symmetry between edges $\V{\xi 1}\leftrightarrow \V{\xi 2}$ with $\xi=1,2$ (and also $\V{1\xi'}\leftrightarrow \V{2\xi'}$ with $\xi'=1,2$) is broken, while edges $\V{\xi 1}\leftrightarrow \V{\xi 2}$ with $\xi=0,1$ (and also $\V{1\xi'}\leftrightarrow \V{2\xi'}$ with $\xi'=0,1$) remain identical. The gate can be accomplished by performing four walks 
\begin{align}\label{eq:Gs:2QB:walk-CZ}
\text{walk 1}:& 
\quad\quad
\V{00} \xrightarrow[]{\text{trivial}} \V{00},
&e^{-i\tau\Lambda}\V{00} = \phantom{+}\V{00},
\\
\text{walk 2}:& 
\quad\quad
\V{01} \xrightarrow[]{{\cal R}_0\text{, via }\V{02}} \V{01},
&e^{-i\tau\Lambda}\V{01} = \phantom{+}\V{01},
\\
\text{walk 3}:& 
\quad\quad
\V{10} \xrightarrow[]{{\cal R}_0\text{, via }\V{20}} \V{10},
&e^{-i\tau\Lambda}\V{10} = \phantom{+} \V{10},
\\
\text{walk 4}:& 
\quad\quad
\V{11} \xrightarrow[]{{\cal R}_\pi\text{, via }\V{12},\V{21},\V{22}} \V{11},
&e^{-i\tau\Lambda}\V{11} = -\V{11},
\end{align}
where ${\cal R}_0$ is a return walk that comes back with no additional phase and ${\cal R}_\pi$ is a return walk that accumulate a phase of $\pi$. Walk 4 in this case probes the reduced symmetry. The subgraphs with non-zero edge amplitudes corresponding to each walk are shown in Fig.~\ref{fig:CZ}. Walk 1 is trivial. Walks 2 and 3 are accomplished by choosing $\tau\Omega_A = 2\pi n_A$, $\tau\Omega_{A'} = 2\pi n_{A'}$ with $n_A,n_{A'}\in \mathbb{Z}$. Walk 4 is performed by choosing

\begin{eqnarray}\label{eq:Gs:2QB:eqs}
\left\{\begin{array}{lcl}
|\Omega_{A'}\Omega^*_B - \Omega_A\Omega^*_{B'}|
= 
\frac{\pi^2}{\tau^2} nm
\\
|\Omega_A\Omega_B + \Omega_{A'}\Omega_{B'}|
=
\frac{\pi^2}{\tau^2}
\frac{nm}{|a|}
\sqrt{(n+m)^2 - (\frac{nm}{|a|}+|a|)^2}
\\
|a| = nm/\sqrt{n_A^2 + n_{A'}^2}
\\
m\le
\sqrt{n_A^2 + n_{A'}^2}
\le n
\end{array}\right.,
\end{eqnarray}
where $n$ and $m$ are any odd non-equal integers ordered as $0<m<n$. This can be obtained by transforming the subgraph corresponding to walk 4, see Fig.~\ref{fig:CZ}, to a linear chain of 4 states as introduced in Sec.~\ref{sec:QWG-1D} and  detailed in Appendix~\ref{sec:SG:romb}. Return walks on the latter graph can be obtained analytically for arbitrary complex hopping amplitudes, see Sec.~5.3 in Ref.~\onlinecite{Solenov-QW}, resulting in system (\ref{eq:Gs:2QB:eqs}).

\section{Summary}

We demonstrated that quantum walks is a general mathematical description naturally encompassing any quantum gates in gate-based quantum computing architectures. It follows from the observation that classical driving field (as the limiting case of quantum control field) connects available states into a complex {\it quantum network} for the duration of the gate. The structure in this network depends on two factors: (i) physical interactions between underlaying qubits and (ii) spectral composition of the control field. More (physically) interacting systems produce less-symmetric networks. Control field can manipulate node-to-node transition rates (weights associated with edges) as allowed per symmetry and can also increase this symmetry. Quantum gates are realized in such Hilbert space network as free evolution. The size of the network depends on spectral composition of control field and is typically very small (two or three connected nodes) for traditionally-designed gates. Yet much larger networks can be easily assembled and were recently shown\cite{solenov4,Solenov-QW} to produce much faster gates under resonant approximation when gate evolution turns into continuous time quantum walks.

We also demonstrated that quantum walk description is applicable to a driven multiqubit system in general, irrespective of approximations made. Evolution of such system  is identical to that of coined quantum walk with one multi-state coin per Fourier harmonic of the control field, as has been summarized in Eq.~(\ref{eq:Gs:walk}). The states of the coin (at each frequency) identify different quasienery bands in the system. They originate from bosonic equidistant energy ladder of quantum field underlaying classical time-dependent control.

The overall evolution of periodically driven system can develop three distinct time scales: (i) fast, (ii) intermediate, and (iii) slow. The fast evolution scale correctly accounts for counter-rotating terms in quantum systems driven by $\cos\omega t$ field, see Eq.~(\ref{eq:QWG:1W:V+}). It does not conserve the total number of excitations (coin + qubits). The intermediate time scale involves evolution which conserves the total number of excitations of the coin-qubits system. The slowest time scale concerns with resonant processes in which some transitions in the qubit system are in exact resonance with coin transitions (at a given frequency), i.e., in Eq.~(\ref{eq:QWG:1W:delta}) $\delta_i^{(-)} = 0$. 

All these time scales are necessarily much faster than the time scale of the overall gate applied to the qubit via the driving field. When time scales (i), (ii) and (iii) are clearly identifiable and are well separated, two consecutive approximations---rotating wave and resonant approximations---can be made. Rotating wave approximation ignores processes at time scale (i), and resonant approximation also ignores processes at time scale (ii). These approximations are naturally identifiable in the quantum walk description---each time scale involves its own quantum walk, which averages to identity if that time scale is well separated (involves many steps of walk without significant interference from other processes). Quantum walks on both scales (i) and (ii) depend on the states of the coin at each frequency, see Eqs~(\ref{eq:QWG:NW:walk}) and (\ref{eq:QWG:RWA:walk}). At time scale (iii), all coins are completely factored out and the evolution is that of a continuous time quantum walk, see Eq.~(\ref{eq:QWG:RA:walk}).

If time scales (i), (ii), and (iii) are not sufficiently distinguishable (corresponding walks are too short), rotating wave and resonant approximations will lead to incorrect description---coins' degrees of freedom will not be accounted correctly (rotating wave approximation) or will be completely ignored (resonant approximation), which effectively traces out all coins resulting in decoherence. While initial system of qubits is a {\it closed quantum system}, classical oscillating driving field performing the gate introduces additional ``hidden'' degrees of freedom---quasienergy bands. These degrees of freedom appear as quantum coins influencing qubit evolution. They provide an additional channel to {\it carry information} between different steps of the evolution and, thus, can be thought of as {\it environmental degrees of freedom}. As the result, rotating wave and resonant approximations in this system can be interpreted as Markovian assumption in which information transfered from qubits to environment (coins) is lost and does not return back to qubits.

In Sec.~\ref{sec:Gs} we demonstrated that the symmetry of the network edges ${\cal S}_i$ determines how evolution is split between scales (iii) and (i) and (ii). As the result, it is not an artifact of the resonant approximation, but rather a fundamental factor in the overall evolution of the system. It is also the key factor in forming entangling quantum gates. Entanglement can only be manipulated if the symmetry is sufficiently low, otherwise only single-qubit control is possible. Examples of several gates were given.

\begin{appendices}

\section{Reducing complex graphs}\label{sec:SG}

Here we give several examples of transformation $G\to G'$ introduced in Eq.~(\ref{eq:QWG-1D:L}) together with transformation of the diagonal energy term, Eq.~(\ref{eq:QWG-1D:H0}), forming graph $G_0$, and producing additional continuous time quantum walk factor with graph $G''$. Only local transformations within subgraphs $\delta G$ and $\delta G_0$ of the graph $G$ and $G_0$  are given. The entire graph can be manipulated by applying multiple transformations in any order as necessary as far as they operate on subgraph $G-G_Q$, that is outside of the qubit domain, or at least in subgraph $G-G_G$ ($G_G\in G_Q$), where $G_G$ is a collection of qubit basis states affected by the gate. Rotation of states in $G_G$, which are the end points (initial and final) of all quantum walks performing the gate, may scramble operation of the gate.

\subsection{One-segment branch}\label{sec:SG:1s}

We first consider a linear subgraph $\delta G$ with a single-segment branch and arbitrary complex hopping amplitudes (edges). This graph can be transformed by moving the branch by two segments along the chain. Applying this local rotations several times, depending on the structure of the rest of the graph, may remove the branch. When several non-connected branches are present, the procedure can be applied iteratively to each. The cases of connected branches (loops) are addressed in the next subsections. The total number of available free parameters (amplitudes) remains unchanged.

\begin{figure}\begin{center}
\includegraphics[width=0.6\textwidth]{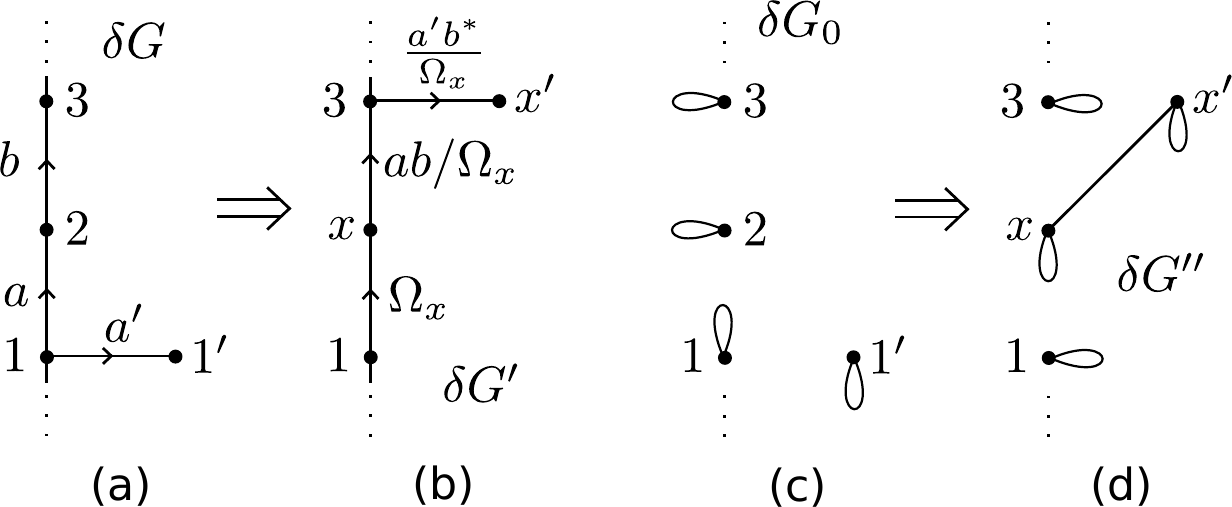}
\vspace*{8pt}
\caption{\label{fig:one-branch}
Transformations  ${\cal L}^T \delta G \to \delta G'$ and ${\cal L}^T \delta G_0 \to \delta G''$ for a linear chain with a single-segment branch.
}\end{center}\end{figure}

The Hamiltonian of subgraph $\delta G$ is
\begin{eqnarray}\label{eq:SG:1s:H_G}
\hat{H}_{\delta G} = a \V{1}\iV{2} + a' \V{1}\iV{1'} + b \V{2}\iV{3} + h.c.,
\end{eqnarray}
as illustrated in Fig.~\ref{fig:one-branch}(a). It is connected to graph $G$ via vertices 1 and 3. The first step in the transformation is to define two orthogonal states $\V{x}$ and $\V{x'}$ as
\begin{eqnarray}\label{eq:SG:1s:xxp}
\left\{\begin{array}{l}
\V{x} = \frac{a^*\V{2} + a'^*\V{1'}}{\Omega_x}
\\
\V{x'} = \frac{a'\V{2} - a\V{1'}}{\Omega_x}
\end{array}\right.,
\end{eqnarray}
with
\begin{eqnarray}\label{eq:1s:Omega_x}
\Omega_x = \sqrt{|a|^2+|a'|^2}.
\end{eqnarray}
This rotation replaces the first two terms in Eq.~(\ref{eq:SG:1s:H_G}) with $\Omega_x\V{1}\iV{x}$. The last term splits into two, because 
\begin{eqnarray}\label{eq:SG:1s:1-2}
\left\{\begin{array}{l}
\V{2} = \frac{a\V{x} + a'^*\V{x'}}{\Omega_x}
\\
\V{1'} = \frac{a'\V{x} - a^*\V{x'}}{\Omega_x}
\end{array}\right..
\end{eqnarray}
As the result, the transformed Hamiltonian is
\begin{eqnarray}\label{eq:SG:1s:H_G'}
\hat{H}_{\delta G'} = \Omega_x \V{1}\iV{x} 
+ \frac{ab}{\Omega_x} \V{x}\iV{3} 
+ \frac{a'b^*}{\Omega_x} \V{3}\iV{x'} 
+ h.c.
\end{eqnarray}
The corresponding graph $\delta G'$ is shown in Fig.~\ref{fig:one-branch}(b). 

The same rotation will transform graph $\delta G_0$, see Fig.~\ref{fig:one-branch}(c), with Hamiltonian
\begin{eqnarray}\label{eq:SG:1s:H_G0}
\hat{H}_{\delta G_0} = E_1\V{1}\iV{1}+E_2\V{2}\iV{2}+E_3\V{3}\iV{3}
+E_{1'}\V{1'}\iV{1'},
\end{eqnarray}
by modifying
\begin{eqnarray}\label{eq:SG:1s:2}
\V{2}\iV{2} = \frac{
|a|^2\V{x}\iV{x}
+|a'|^2\V{x'}\iV{x'}
+aa'\V{x}\iV{x'}
+a'^*a^*\V{x'}\iV{x}
}{\Omega_x^2},
\\\label{eq:SG:1s:1'}
\V{1'}\iV{1'} = \frac{
|a'|^2\V{x}\iV{x}
+|a|^2\V{x'}\iV{x'}
-a'a\V{x}\iV{x'}
-a^*a'^*\V{x'}\iV{x}
}{\Omega_x^2} .
\end{eqnarray}
As the result, we obtain
\begin{eqnarray}\label{eq:SG:1s:H_G0}
&&\hat{H}_{\delta G''} = E_1\V{1}\iV{1}+E_3\V{3}\iV{3}
+\frac{E_2|a|^2+E_{1'}|a'|^2}{\Omega_x^2}
\V{x}\iV{x}
\\\nonumber
&&+\frac{E_2|a'|^2+E_{1'}|a|^2}{\Omega_x^2}
\V{x'}\iV{x'}
+\frac{E_2-E_{1'}}{\Omega_x^2}
(aa'\V{x}\iV{x'} + h.c.),
\end{eqnarray}
as shown in Fig.~\ref{fig:one-branch}(d).

\subsection{square four-segment loop}\label{sec:SG:4l}

\begin{figure}\begin{center}
\includegraphics[width=0.9\textwidth]{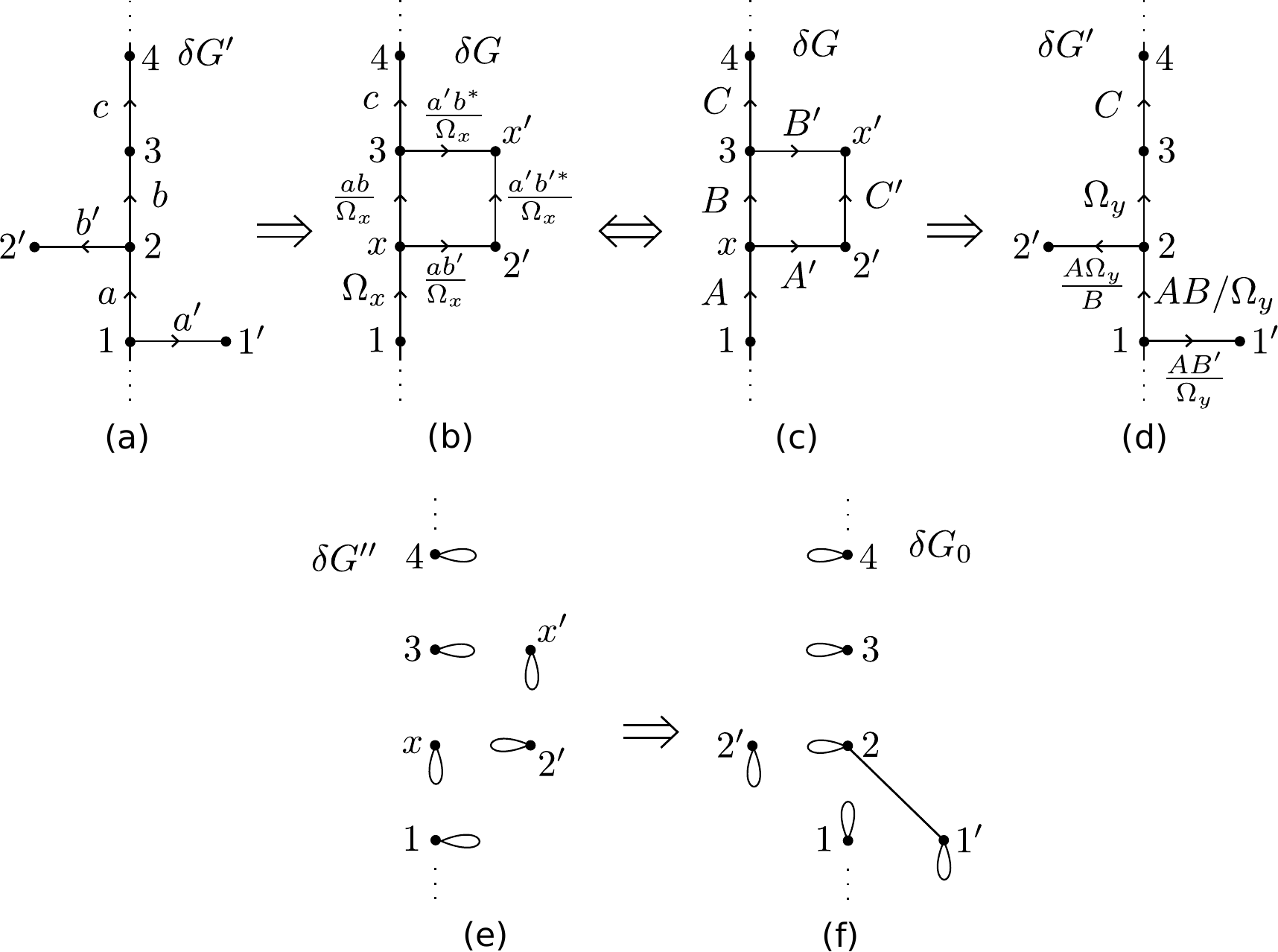}
\caption{\label{fig:4-loop}
Transformations  ${\cal L}^T \delta G \leftrightarrow \delta G'$ and ${\cal L}^T \delta G_0 \leftrightarrow \delta G''$ connecting a linear chain segment with two single-segment branches and a linear chain segment with a square loop.
}
\end{center}\end{figure}

Here we derive the transformation connecting $\delta G$ that is a four-segment loop on a chain (two connected branches), see Figs.~\ref{fig:4-loop}(b) and (c), with graph $\delta G'$ that is a chain with two single segment branches next to each other see Figs.~\ref{fig:4-loop}(a) and (d). We start with the inverse transformation, i.e., $\delta G'\to\delta G$. The initial Hamiltonian is
\begin{eqnarray}\label{eq:SG:4l:H_G'}
\hat{H}_{\delta G'} = a\V{1}\iV{2} + a'\V{1}\iV{1'}
+ b\V{2}\iV{3} + b'\V{2}\iV{2'}
+c\V{3}\iV{4} + h.c.
\end{eqnarray}
The transformation (\ref{eq:SG:1s:xxp}) can be applied to states $\V{2}$ and $\V{1'}$ as before. Now, however, state $\V{2}$ is also connected to $\V{2'}$ and due to Eq.~(\ref{eq:SG:1s:1-2}) we obtain
\begin{eqnarray}\label{eq:SG:4l:H14:new}
\hat{H}_{\delta G}\to
\Omega_x\V{1}\iV{x} 
\!+\!
\frac{ab}{\Omega_x}\V{x}\iV{3}
\!+\!
\frac{b^*a'\!\!}{\Omega_x}\V{3}\iV{x'} 
\!+\!
\frac{ab'\!\!}{\Omega_x}\V{x}\iV{2'} 
\!+\!
\frac{b'a'^*\!\!\!}{\Omega_x}\V{x'}\iV{2'} 
\!+\!
c\V{3}\iV{4} 
\!+\!
h.c.
\end{eqnarray}
The corresponding graph, shown in Fig.~\ref{fig:4-loop}(b), is a four-segment loop attached to a linear chain.
The transformation $\delta G\to\delta G'$ can be found by solving system
\begin{eqnarray}\label{eq:SG:4l:rel}
\left\{\begin{array}{rcl}
C &=& c
\\
A &=& \Omega_x
\\
A' &=& \frac{ab'}{\Omega_x}
\\
B &=& \frac{ab}{\Omega_x}
\\
B' &=& \frac{a'b^*}{\Omega_x}
\\
C' &=& \frac{a'b'^*}{\Omega_x}
\end{array}\right.,
\end{eqnarray}
as follows from Figs.~\ref{fig:4-loop}(b) and (c).
The loop graph [see Fig.~\ref{fig:4-loop}(c)] has one extra complex parameter as compared to graph with two branches in Figs.~\ref{fig:4-loop}(a) and (d). Therefore transformation from a loop graph to a graph with two branches [see Figs.~\ref{fig:4-loop}(c) and (d)] can only occur if the number of free parameters is reduced. The necessary condition  can be derived by comparing equations in system (\ref{eq:SG:4l:rel}). This yields
\begin{eqnarray}\label{eq:SG:4l:cond}
\frac{B'}{B^*}
=
\frac{C'}{A'^*}.
\end{eqnarray}
The rest of the solution is
\begin{eqnarray}\label{eq:SG:4l:abc}
\left\{\begin{array}{rcl}
a &=& \frac{AB}{\Omega_y}
\\
a' &=& \frac{AB'}{\Omega_y}
\\
b &=& \Omega_y = \sqrt{|B|^2+|B'|^2}
\\
b' &=& \frac{A'}{B}\Omega_y
\\
c&=&C
\end{array}\right. .
\end{eqnarray}
The resulting two single-segment branches [see Fig.~\ref{fig:4-loop}(d)] can be moved to one of the ends of the linear chain, one after the other, by iteratively applying transformation derived in Sec.~\ref{sec:SG:1s}.

Under this transformation the diagonal $\delta G_0$ graph with Hamiltonian
\begin{eqnarray}\label{eq:SG:4l:H_G0}
\hat{H}_{\delta G_0} = E_1\V{1}\iV{1}+E_x\V{x}\iV{x}+E_3\V{3}\iV{3}+E_4\V{4}\iV{4}
+E_{x'}\V{x'}\iV{x'}+E_{2'}\V{2'}\iV{2'},
\end{eqnarray}
shown in Fig.~\ref{fig:4-loop}(e) is adjusted via
\begin{eqnarray}\label{eq:SG:4l:2}
\V{x}\iV{x} = \frac{
|a|^2\V{2}\iV{2}
+|a'|^2\V{1'}\iV{1'}
+a^*a'\V{2}\iV{1'}
+a'^*a\V{1'}\iV{2}
}{\Omega_x^2},
\\\label{eq:SG:4l:1'}
\V{x'}\iV{x'} = \frac{
|a'|^2\V{2}\iV{2}
+|a|^2\V{1'}\iV{1'}
-a'a^*\V{2}\iV{1'}
-aa'^*\V{1'}\iV{2}
}{\Omega_x^2} ,
\end{eqnarray}
producing graph $\delta G''$ with Hamiltonian
\begin{eqnarray}\label{eq:SG:4l:H_G0}
\hat{H}_{\delta G''} &=& E_1\V{1}\iV{1}+E_3\V{3}\iV{3}+E_4\V{4}\iV{4}+E_{2'}\V{2'}\iV{2'}
\\\nonumber
&&
+\frac{E_x|a|^2+E_{x'}|a'|^2}{\Omega_x^2}
\V{2}\iV{2}
+\frac{E_x|a'|^2+E_{x'}|a|^2}{\Omega_x^2}
\V{1'}\iV{1'}
\\
&&+\frac{E_x-E_{x'}}{\Omega_x^2}
(a'a^*\V{2}\iV{1'} + h.c.),
\end{eqnarray}
shown in Fig.~\ref{fig:4-loop}(f).

\subsection{diagonal square loop}\label{sec:SG:romb}

\begin{figure}\begin{center}
\includegraphics[width=0.7\textwidth]{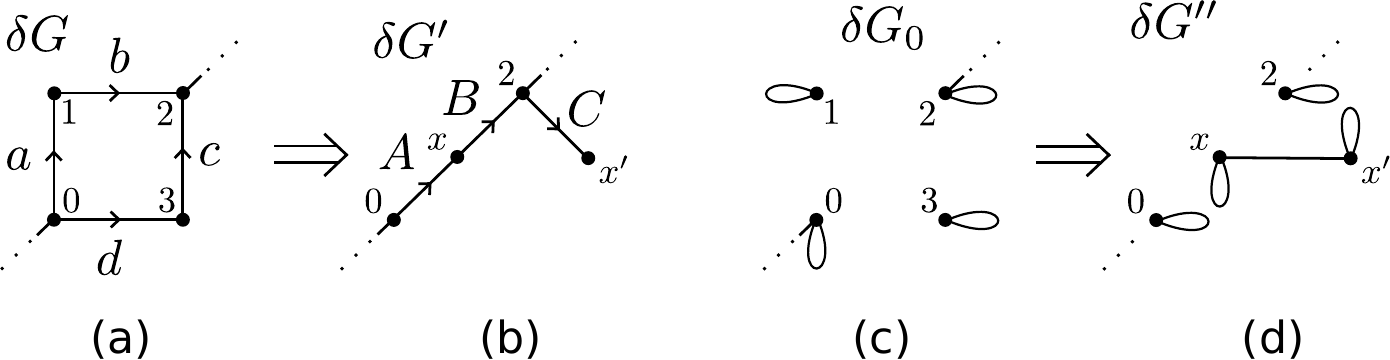}
\caption{\label{fig:romb}
The transformation between four-segment loop, graph $\delta G$, inserted via its diagonal and a single-branch graph $\delta G'$.
}
\end{center}\end{figure}

Here we transform a square loop attached to the rest of the graph via its diagonal. The Hamiltonian of the four-segment square part, graph $\delta G$, shown in Fig.~\ref{fig:romb}(a), is
\begin{eqnarray}\label{eq:SG:romb:H02}
\hat{H}_{\delta G} = a \V{0}\iV{1} + b \V{1}\iV{2} 
+ c \V{3}\iV{2} + d \V{0}\iV{3} + h.c.
\end{eqnarray}
The first and the fourth terms can be combined to define two new orthogonal states $\V{x}$ and $\V{x'}$
\begin{eqnarray}\label{eq:SG:romb:xxp}
&&\left\{\begin{array}{lcl}
\V{x}&=&\frac{a^*\V{1} + d^*\V{3}}{\Omega_x}
\\
\V{x'}&=&\frac{d\V{1} - a\V{3}}{\Omega_x}
\end{array}\right.,
\\\label{eq:romb:Omega_x}
&&\quad\,\,
\Omega_x = \sqrt{|a|^2+|d|^2}.
\end{eqnarray}
The remaining two terms, i.e. the second and the third, are transformed into the new basis
\begin{eqnarray}\label{eq:SG:romb:12}
b\V{1}\iV{2}&\to&
 \frac{ab}{\Omega_x}\V{x}\iV{2}
+ \frac{bd^*}{\Omega_x}\V{x'}\iV{2},
\\\label{eq:romb:32}
c\V{3}\iV{2}&\to&
 \frac{cd}{\Omega_x}\V{x}\iV{2}
- \frac{a^*c}{\Omega_x}\V{x'}\iV{2}.
\end{eqnarray}
As the result, we obtain
\begin{eqnarray}\label{eq:romb:H02new}
\hat{H}_{\delta G'} = A \V{0}\iV{x} + B \V{x}\iV{2} 
+ C \V{2}\iV{x'} + h.c.,
\end{eqnarray}
with graph $\delta G'$ shown in Fig.~\ref{fig:romb}(b), where
\begin{eqnarray}\label{eq:SG:romb:ABC}
\left\{\begin{array}{rcl}
A&=& \Omega_x
\\
B&=& \frac{ab+cd}{\Omega_x}
\\
C&=& \frac{b^*d-ac^*}{\Omega_x}
\end{array}\right. .
\end{eqnarray}
Note that the number of parameters (graph edges) is reduced from four to three complex numbers.

Under this transformation the diagonal $\delta G_0$ graph with Hamiltonian
\begin{eqnarray}\label{eq:SG:romb:H_G0}
\hat{H}_{\delta G_0} = E_0\V{0}\iV{0}+E_1\V{1}\iV{1}+E_2\V{2}\iV{2}+E_3\V{3}\iV{3},
\end{eqnarray}
shown in Fig.~\ref{fig:romb}(c) is transformed to
\begin{eqnarray}\label{eq:SG:romb:H_G0}
\hat{H}_{\delta G''} &=& 
E_0\V{0}\iV{0}+E_2\V{2}\iV{2}
\\\nonumber
&&
+\frac{E_1|a|^2+E_3|d|^2}{\Omega_x^2}
\V{x}\iV{x}
+\frac{E_1|d|^2+E_3|a|^2}{\Omega_x^2}
\V{x'}\iV{x'}
\\
&&+\frac{E_1-E_3}{\Omega_x^2}
(ad\V{x}\iV{x'} + h.c.),
\end{eqnarray}
as shown in Fig.~\ref{fig:romb}(d).

\subsection{Six-segment loop inserted via diagonal}\label{sec:SG:2s}

\begin{figure}\begin{center}
\includegraphics[width=0.8\textwidth]{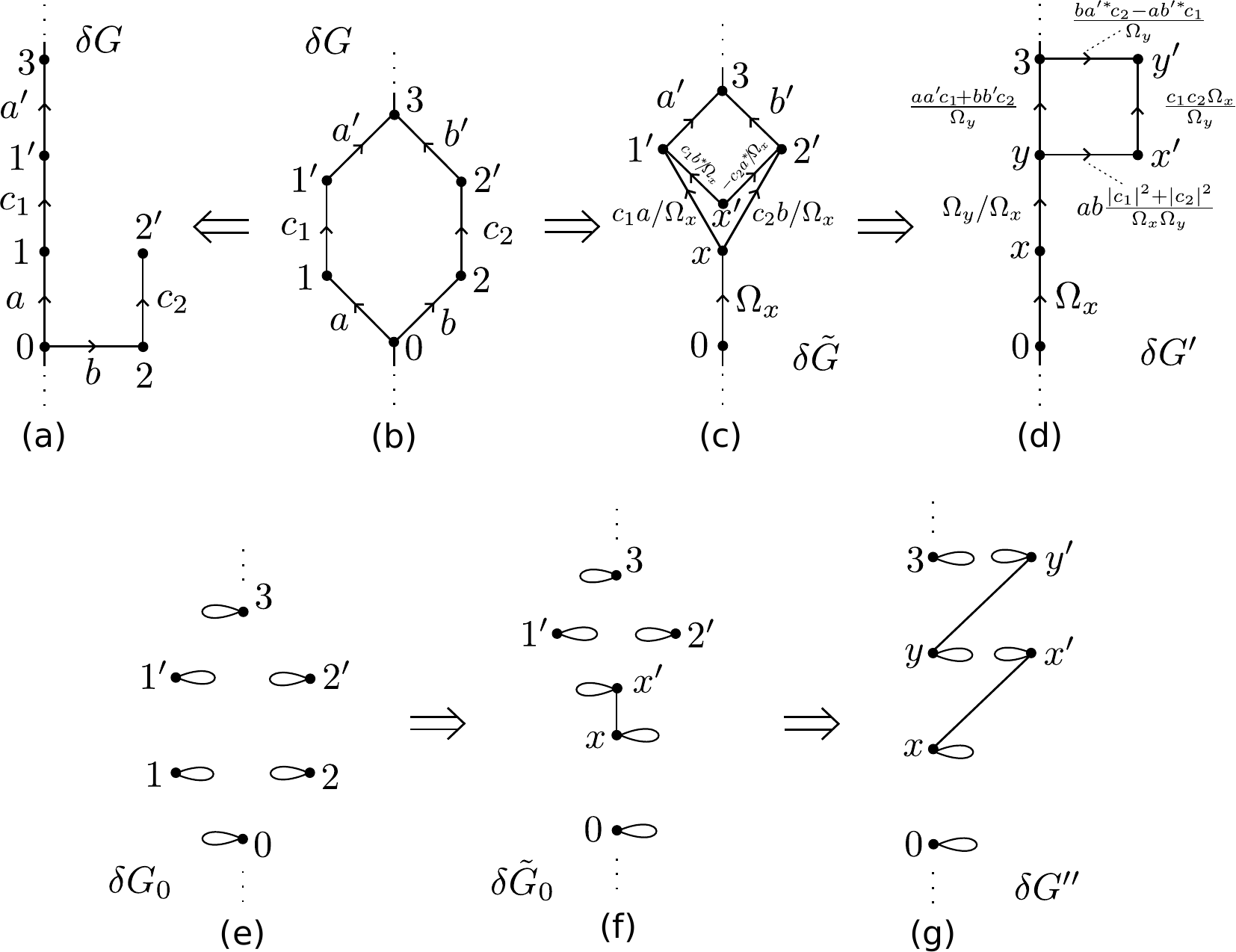}
\caption{\label{fig:diamond}
The transformation between a six-segment loop breaking a linear chain graph and a four-segment edge-sharing loop, with a linear chain with a two-segment branch as a special case.
}
\end{center}\end{figure}

Here we demonstrate how to transform a six-segment loop subgraph inserted via its (largest) diagonal, see Fig.~\ref{fig:diamond}(b), into a four-segment square loop subgraph, see Fig.~\ref{fig:diamond}(d), discussed in the previous subsection. In the specific case of $b'=0$, it also demonstrates how to reduce a linear chain with a two-segment branch, Fig.~\ref{fig:diamond}(a). The Hamiltonian corresponding to the subgraphs $\delta G$, shown in Fig.~\ref{fig:diamond}(a) and (b), is
\begin{eqnarray}\label{eq:SG:2s:H03}
\hat{H}_{\delta G} = a \V{0}\iV{1} + b \V{0}\iV{2} 
+ c_1 \V{1}\iV{1'} + c_2 \V{2}\iV{2'}
+ a' \V{1'}\iV{3} + b' \V{2'}\iV{3}
+ h.c.
\end{eqnarray}
As before, the first two terms can be reduced by rotating the states $\V{1}$ and $\V{2}$ to introduce new orthogonal states
\begin{eqnarray}\label{eq:SG:2s:xxp}
\left\{\begin{array}{rcl}
\V{x} &=& \frac{a^*\V{1} + b^*\V{2}}{\Omega_x}
\\
\V{x'} &=& \frac{b\V{1} - a\V{2}}{\Omega_x}
\end{array}\right.,
\\\label{eq:SG:2s:Omega_x}
\Omega_x = \sqrt{|a|^2 + |b|^2}.
\end{eqnarray}
This procedure transforms $\delta G$ into $\delta \tilde G$ described by Hamiltonian
\begin{eqnarray}\nonumber
\hat{H}_{\delta \tilde G} &=& \Omega_x \V{0}\iV{x} 
+ \V{x}\frac{ac_1\iV{1'} + bc_2\iV{2'}}{\Omega_x} 
+ \V{x'}\frac{b^*c_1\iV{1'} - a^*c_2\iV{2'}}{\Omega_x} 
\\\label{eq:SG:2s:H03aux}
&&+ a' \V{1'}\iV{3} + b' \V{2'}\iV{3}
+ h.c.,
\end{eqnarray}
shown in Fig.~\ref{fig:diamond}(c). Further binary rotation involving states $\V{1'}$ and $\V{2'}$
\begin{eqnarray}\label{eq:SG:2s:yyp}
&&\left\{\begin{array}{rcl}
\V{y} &=& \frac{a^*c_1^*\V{1'} + b^*c_2^*\V{2'}}{\Omega_y}
\\
\V{y'} &=& \frac{bc_2\V{1'} - ac_1\V{2'}}{\Omega_x}
\end{array}\right.,
\\\label{eq:2s:Omega_y}
&&\quad\quad\,\Omega_y = \sqrt{|ac_1|^2 + |bc_2|^2},
\end{eqnarray}
simplifies the second term in Eq.~(\ref{eq:SG:2s:H03aux}), yielding
\begin{eqnarray}\nonumber
&&\hat{H}_{\delta G'} = \Omega_x \V{0}\iV{x} 
+ \frac{\Omega_y}{\Omega_x}\V{x}\iV{y} 
+ ab\frac{|c_1|^2-|c_2|^2}{\Omega_x\Omega_y}\V{y}\iV{x'}
\\\label{eq:SG:2s:H03new}
&&+ \frac{c_1c_2\Omega_x}{\Omega_y}\V{x'}\iV{y'}
+ \frac{aa'c_1+bb'c_2}{\Omega_y}\V{y}\iV{3}
+ \frac{ba'^*c_2-ab'^*c_1^*}{\Omega_y}\V{3}\iV{y'}
+ h.c.
\end{eqnarray}
This produces graph $\delta G'$ shown in Fig.~\ref{fig:diamond}(d).

The corresponding graph $\delta G_0$, described by Hamiltonian
\begin{eqnarray}\label{eq:SG:2s:H_G0}
\hat{H}_{\delta G_0} = E_0\V{0}\iV{0}+E_1\V{1}\iV{1}+E_2\V{2}\iV{2}+E_3\V{3}\iV{3}
+E_{1'}\V{1'}\iV{1'}+E_{2'}\V{2'}\iV{2'}
\end{eqnarray}
and shown in Fig.~\ref{fig:diamond}(e), is first transformed to $\delta \tilde G_0$ using Eq.~(\ref{eq:SG:1s:2}) with replacements $a'\to b$ and $\V{1'}\to \V{2}$. The intermediate graph Hamiltonian is
\begin{eqnarray}\label{eq:SG:2s:H_tildeG0}
\hat{H}_{\delta \tilde G_0} &=& 
E_0\V{0}\iV{0}+E_3\V{3}\iV{3}+E_{1'}\V{1'}\iV{1'}+E_{2'}\V{2'}\iV{2'}
\\\nonumber
&+&\frac{E_2|a|^2+E_{1}|b|^2}{\Omega_x^2}
\V{x}\iV{x}
+\frac{E_2|b|^2+E_{1}|a|^2}{\Omega_x^2}
\V{x'}\iV{x'}
+\frac{E_2-E_{1}}{\Omega_x^2}
(ab\V{x}\iV{x'} + h.c.),
\end{eqnarray}
as shown in Fig.~\ref{fig:diamond}(f). The second rotation, Eq.~(\ref{eq:SG:2s:yyp}), transforms $E_{1'}\V{1'}\iV{1'}+E_{2'}\V{2'}\iV{2'}$. Finally, we obtain the Hamiltonian describing graph $\delta G''$
\begin{eqnarray}\label{eq:SG:2s:H_G''}
\hat{H}_{\delta G''} &=& 
E_0\V{0}\iV{0}+E_3\V{3}\iV{3}
\\\nonumber
&+&\frac{E_{1'}|ac_1|^2+E_{2'}|bc_2|^2}{\Omega_y^2}
\V{y}\iV{y}
+\frac{E_{1'}|bc_2|^2+E_{2'}|ac_1|^2}{\Omega_y^2}
\V{y'}\iV{y'}
\\\nonumber
&+&\frac{E_{1'}-E_{2'}}{\Omega_y^2}
(abc_1c_2\V{y}\iV{y'} + h.c.)
\\\nonumber
&+&\frac{E_2|a|^2+E_{1}|b|^2}{\Omega_x^2}
\V{x}\iV{x}
+\frac{E_2|b|^2+E_{1}|a|^2}{\Omega_x^2}
\V{x'}\iV{x'}
+\frac{E_2-E_{1}}{\Omega_x^2}
(ab\V{x}\iV{x'} + h.c.),
\end{eqnarray}
shown in Fig.~\ref{fig:diamond}(g).

\end{appendices}


\end{document}